\documentclass[preprint,aps,floatfix,nofootinbib,a4paper,superscriptaddress]{revtex4-1}

\usepackage{amsmath, amssymb, amsfonts, amsthm, latexsym, epsfig, mathrsfs, xcolor, bbm, slashed}

\usepackage[inline]{enumitem}

\usepackage{setspace}
\usepackage[marginal, multiple]{footmisc}

\usepackage[utf8]{inputenc}

\usepackage[colorlinks, allcolors=blue!70!black, linktocpage]{hyperref}

\numberwithin{equation}{section}

\usepackage{cleveref}

\usepackage{microtype}

\usepackage{floatrow}
\usepackage{cancel}

\let\OLDthebibliography\thebibliography
\renewcommand\thebibliography[1]{%
	\setstretch{1.079} 
	\OLDthebibliography{#1}%
	\small %
	\setlength{\itemsep}{0.2\baselineskip} 
}

\let\OLDfootnote\footnote
\renewcommand\footnote[1]{%
	\setlength{\footnotesep}{0.75\baselineskip}%
    \count\footins = 750%
	{\footnotesize \OLDfootnote{#1}}%
}

\setlist[enumerate]{noitemsep, label=(\arabic*), ref=(\arabic*)}

\newlist{condlist}{enumerate}{2}
\setlist[condlist,1]{noitemsep, label=(\arabic*), ref=(\arabic*)}
\setlist[condlist,2]{noitemsep, label=(\alph*), ref=(\arabic{condlisti}.\alph*)}
\crefname{condlisti}{condition}{conditions}
\crefname{condlistii}{condition}{conditions}

\renewcommand\thesection{\arabic{section}}
\renewcommand\thesubsection{\arabic{subsection}}

\makeatletter
\def\p@subsection{\thesection.}
\def\p@subsubsection{\thesection.\thesubsection.}
\makeatother

\theoremstyle{plain}

\theoremstyle{definition}
\newtheorem{definition}{Definition}

\theoremstyle{remark}
\newtheorem{remark}{Remark}[section]

\crefname{equation}{Eq.}{Eqs.}
\creflabelformat{equation}{#2#1#3}

\crefname{section}{\S}{\S}
\crefname{appendix}{Appendix}{Appendices}
\crefname{figure}{Fig.}{Figs.}

\crefname{definition}{Def.}{Defs.}
\crefname{prop}{Prop.}{Props.}
\crefname{lemma}{Lemma}{Lemmas}
\crefname{corollary}{Cor.}{Cors.}
\crefname{thm}{Theorem}{Theorems}
\crefname{remark}{Remark}{Remarks}

\crefname{ass}{Assumptions}{Assumptions}
\crefname{property}{Properties}{Properties}

\newcommand{\be}{\begin{equation}}
\newcommand{\ee}{\end{equation}}

\newcommand{\lb}{\left}
\newcommand{\rb}{\right}

\newcommand{\mc}{\mathcal}

\newcommand{\ms}{\mathscr}
\newcommand{\mf}{\mathfrak}
\newcommand{\bb}{\mathbb}
\renewcommand{\=}{\hateq}

\newcommand{\eqsp}{\, ,\quad} 

\renewcommand{\bar}{\overline}

\newcommand{\half}{\frac{1}{2}}

\newcommand{\abs}[1]{\lb\vert\, #1 \,\rb\vert}		

\newcommand{\Lie}{\pounds} 
\newcommand{\defn}{\mathrel{\mathop:}=} 
\newcommand{\grad}{\nabla}
\newcommand{\del}{\partial}
\renewcommand{\b}{\mathfrak{b}}

\newcommand{\s}{\mathfrak{s}}

\newcommand{\scri}{\ms I}

\newcommand{\hateq}{\mathrel{\mathop {\widehat=} }} 

\newcommand{\floor}[1]{\lb\lfloor{#1}\rb\rfloor} 

\newcommand{\af}[1]{\mathring{#1}} 

\newcommand{\pb}[1]{\underleftarrow{#1}} 

\newcommand{\arctanh}{{\rm arctanh}}

\begin{document}

\setstretch{1.2}


\title{BMS-like symmetries in cosmology}

\author{B\'eatrice Bonga}\email{bbonga@science.ru.nl}
\affiliation{Institute for Mathematics, Astrophysics and Particle Physics, Radboud University, 6525 AJ Nijmegen, The Netherlands}

\author{Kartik Prabhu}\email{kartikprabhu@ucsb.edu}
\affiliation{Department of Physics, University of California, Santa Barbara, CA 93106, USA}

\begin{abstract}
Null infinity in asymptotically flat spacetimes posses a rich mathematical structure; including the Bondi-Metzner-Sachs (BMS) group and the Bondi news tensor that allow one to study gravitational radiation rigorously. However, Friedmann-Lema\^itre-Robertson-Walker (FLRW) spacetimes are not asymptotically flat because their stress-energy tensor does not decay sufficiently fast, and in fact diverges, at null infinity. This class includes matter- and radiation-dominated FLRW spacetimes. We define a class of spacetimes whose structure at null infinity is similar to FLRW spacetimes: the stress-energy tensor is allowed to diverge and the conformal factor is not smooth at null infinity. Interestingly, for this larger class of spacetimes, the asymptotic symmetry algebra is similar to the BMS algebra but not isomorphic to it. In particular, the symmetry algebra is the semi-direct sum of supertranslations and the Lorentz algebra, but it does not have any preferred translation subalgebra. Future applications include studying gravitational radiation in FLRW the full nonlinear theory, including the cosmological memory effect, and also asymptotic charges in this framework.
\end{abstract}

\maketitle
\tableofcontents

\section{Introduction}\label{sec:intro}

For asymptotically flat spacetimes describing isolated systems in vacuum general relativity, it is well-known that the asymptotic symmetries at null infinity are given by the infinite-dimensional Bondi-Metzner-Sachs (BMS) group \cite{BBM,Sachs2,Geroch-asymp}, which is a semi-direct product of the infinite-dimensional group of supertranslations and the Lorentz group. The supertranslations are essentially ``angle-dependent'' translations which are not exact symmetries of flat Minkowski spacetime, but only arise in the asymptotic regime at null infinity and are intimately tied to the presence of gravitational radiation. This ``infinite enhancement'' of the Poincar\'e group of symmetries to BMS symmetries in the presence of gravitational radiation is related to many non-trivial phenomena at null infinity. For example, the gravitational memory effect --- a permanent displacement of test bodies after the passage of a gravitational wave --- can be related to the non-trivial supertranslations at null infinity \cite{He:2014laa,Strominger:2014pwa,Pasterski:2015tva,HIW,Pate:2017fgt,Chatterjee:2017zeb}. This memory effect has been forecasted to be observable by advanced gravitational wave detection methods \cite{Lasky:2016knh,Favata:2009ii,Wang:2014zls,Arzoumanian:2015cxr,Nichols:2017rqr}. Further, associated with the BMS Lie algebra are an infinite number of charges and fluxes \cite{BBM, Sachs1, Sachs2, Penrose, Geroch-asymp, GW, Ashtekar:1981bq, WZ}. These charges and fluxes have been related to the soft graviton theorems \cite{He:2014laa,Strominger:2014pwa,Avery:2015gxa,Campiglia:2015kxa},
and potentially black hole information loss \cite{Hawking:2016msc,Strominger:2017aeh,Strominger:2017zoo}. BMS-like symmetries have also been found on null surfaces in finite regions of the spacetime \cite{Chandrasekaran:2018aop,Chandrasekaran:2019ewn} including black hole horizons \cite{Donnay:2015abr} and cosmological horizons \cite{Donnay:2019zif}, and their connection to memory effects across these horizons have been studied in \cite{Donnay:2018ckb,Rahman:2019bmk}.\\

Another class of spacetimes which also have a future null boundary ``at infinity'' are expanding, spatially flat Friedmann-Lema\^itre-Robertson-Walker (FLRW) spacetimes whose expansion decelerates towards the future. Examples of such spacetimes describe radiation-dominated and matter-dominated cosmologies. The asymptotic symmetries in FLRW spacetimes have recently been studied by Kehagias and Riotto in \cite{KR}. Since the FLRW spacetimes are conformal to Minkowski spacetime, Kehagias and Riotto study the asymptotic behaviour of FLRW spacetimes in a Bondi coordinate system adapted to Minkowski spacetime. While this method of analysis is certainly valid, it obscures the key differences between the asymptotic behaviour of FLRW and Minkowski spacetimes. For example, while the components of the stress-energy tensor of the FLRW spacetime in the Bondi-Sachs coordinate system adapted to the Minkowski spacetime do fall off asymptotically, as shown in \cite{KR}, their falloff is in fact too slow for the stress-energy tensor to even have a finite limit to null infinity (see \cref{rem:coordinate-fall-off})! A more serious error in \cite{KR} is that the asymptotic symmetries of the FLRW spacetime are defined through coordinate transformations of the Bondi-Sachs coordinate system adapted to the Minkowski spacetime. Using this procedure Kehagias and Riotto obtained the BMS algebra as the asymptotic symmetry algebra even for FLRW spacetimes. However, general diffeomorphisms of FLRW spacetimes \emph{cannot} be written as the scale factor times a diffeomorphism of the Minkowski spacetime; one also needs to transform the scale factor (see \cref{rem:KR-symm}). Thus, the symmetry algebra obtained by Kehagias and Riotto does not arise from diffeomorphisms of the FLRW spacetime and should not be considered as the asymptotic symmetry algebra in FLRW spacetimes.\\

The goal of this paper is to reanalyze the asymptotic symmetries at null infinity in FLRW spacetimes using the covariant formalism of a conformal completion \`a la Penrose.  It is well-known that for decelerating FLRW spacetimes the conformal completion has a smooth null boundary \(\scri\) denoting \emph{null infinity} \cite{Carroll-book,Harada:2018ikn}. We investigate the structure at \(\scri\) in FLRW spacetimes in detail and show that while the conformal completion of these FLRW spacetimes looks superficially similar to that of Minkowski there are some crucial differences. Since the FLRW spacetimes are homogenous, the matter stress-energy tensor does not fall off towards null infinity, and in fact, diverges in the limit to \(\scri\) (see \cref{eq:T-exp}). Further, the conformal factor \(\Omega\) relating the physical FLRW spacetime to its conformal completion is not smooth at null infinity and vanishes faster compared to its behaviour in Minkowski spacetime in the sense that the derivative of \(\Omega\) also vanishes at \(\scri\). This behaviour can be captured in a single parameter, denoted by \(s\), which is directly related to the equation of state parameter of the perfect fluid matter in the FLRW spacetime (see \cref{eq:s-w-q}). For decelerating FLRW spacetimes this parameter satisfies \(0 \leq s < 1\) with the case \(s=0\) corresponding to the Minkowski spacetime.

We then define a class of spacetimes that have a \emph{cosmological null asymptote} at infinity. Just as asymptotically flat spacetimes are defined so that their behaviour at null infinity is modeled on that of exact Minkowski spacetime, the spacetimes with a cosmological asymptote are defined so that their behaviour at null infinity is similar to that of decelerating FLRW spacetimes. In particular, the conformal factor in such spacetimes is allowed to be non-smooth and the stress-energy tensor is allowed to diverge at null infinity, parameterized by a number \(s\) as described above for exact FLRW spacetimes (see \cref{def:dna} for details).

Within this class of spacetimes for each \(s\), we derive the \emph{asymptotic symmetry algebra}, denoted by \(\b_s\), which is generated by infinitesimal diffeomorphisms that preserve the asymptotic structure at null infinity. We show that the algebra \(\b_s\) is similar to the BMS algebra --- it is the semi-direct sum of an infinite-dimensional abelian subalgebra of \emph{supertranslations} with the Lorentz algebra. However, since these spacetimes are not asymptotically flat (unless \(s=0\)) this algebra is not isomorphic to the BMS algebra. In particular, we show that there is no longer a preferred translation subalgebra in \(\b_s\), contrary to the structure of the BMS algebra. Rather, the asymptotic symmetry algebra \(\b_s\) is isomorphic to the Lie algebra of the conformal Carroll groups studied in \cite{Duval:2014uva,Duval:2014lpa,Hartong:2015xda,Ciambelli:2019lap}. \\

The rest of this paper is organized as follows. In \cref{sec:conformalcompletionFLRW}, we discuss in detail the conformal completion of FLRW spacetimes making explicit the similarities and differences with the conformal completion of Minkowski spacetime. In \cref{sec:dna}, we present the definition of the class of spacetimes with a cosmological null asymptote which behave like an FLRW spacetime near null infinity, work out the consequences of the Einstein equation and demonstrate that the class of spacetimes with a cosmological null asymptote is at least as big as the class of asymptotically flat spacetimes. In \cref{sec:symm}, we discuss the universal structure and asymptotic symmetry algebra. Given the widespread use of Bondi-Sachs coordinates and their convenience in explicit calculations, in \cref{sec:BS}, we construct from the geometric definition two types of Bondi-Sachs coordinates that we believe might be useful in future studies of this class of spacetimes. We conclude in \cref{sec:dis} and discuss different applications of this framework. In \cref{sec:trans-ideal}, we show that the asymptotic symmetry algebras \(\b_s\) do not have a preferred subalgebra of translations unless \(s=0\), in which case the algebra is isomorphic to the BMS algebra of asymptotically flat spacetimes. In \cref{sec:open} we briefly summarize the conformal completions of spatially open FLRW spacetimes and show that their behaviour at null infinity is very different from the spatially flat case considered in the main paper.\\

Our conventions are as follows. The spacetime is \(4\)-dimensional with a metric of signature \((-,+,+,+)\). We use abstract indices \(a,b,c,\ldots\) to denote tensor fields. We will also use indices \(A,B,C,\ldots\) to denote tensor components in some choice of coordinate system on a \(2\)-sphere. Quantities defined on the physical spacetime will be denoted by a ``hat'', while the ones on the conformally-completed unphysical spacetime are without the ``hat'', e.g. \(\hat g_{ab}\) is the physical metric while \(g_{ab}\) is the unphysical metric on the conformal-completion. The symbol \(\hateq\) will be used to denote equality when evaluated at points of null infinity \(\scri\). The rest of our conventions follow those of Wald \cite{Wald-book}.

\section{Conformal completion of decelerating FLRW spacetimes}
\label{sec:conformalcompletionFLRW}

In this section, we discuss in detail the conformal completion of decelerating FLRW spacetimes to motivate the conditions of the class of spacetimes defined in the next section. We will focus on the case of spatially flat FLRW spacetimes --- spatially closed FLRW spacetimes do not have a future null infinity and are not of interest in this paper, and while spatially open FLRW spacetimes have a future null infinity, their behavior is very distinct from the spatially flat case (for details, see \cref{sec:open}). 

In particular, we highlight the key differences between the conformal completion of decelerating FLRW spacetimes and asymptotically flat spacetimes that arise due to the presence of homogenous and isotropic matter.\\

The physical metric $\hat{g}_{ab}$ of decelerating, spatially flat FLRW spacetimes is described by the line element\footnote{In this section we use \(r\) for the radial coordinate in the Minkowski spacetime. This should not be confused with the Bondi-Sachs coordinate defined in \cref{sec:BS-phys}.}
\begin{equation}\label{eq:FLRW}
    d\hat s^2 = a^2(\eta) \left( - d\eta^2 + dr^2 + r^2 S_{AB} dx^A dx^B \right) \quad \text{with } a(\eta) = \left(\frac{\eta}{\eta_0} \right)^{s/(1-s)} \; ,
\end{equation}
where we have used the \emph{conformal time} coordinate \(\eta \in (0,\infty)\), the radial coordinate $r \in [0, \infty)$, and \(x^A\) are some coordinates with \(S_{AB}\) being the unit round metric on the \(2\)-sphere \(\bb S^2\). The function \(a(\eta)\) is the \emph{scale factor} which describes the expansion of the universe with time,\footnote{Reversing the direction of time gives us a contracting FLRW universe which can be analyzed in the same manner.} and \(\eta_0\) is a normalization constant so that \(a(\eta=\eta_0) = 1\). The parameter \(s\) is related to the stress-energy as described below (see \cref{eq:s-w-q}), and is introduced for later convenience.

The FLRW spacetimes satisfy the Einstein equation
\be
    \hat G_{ab} = 8\pi \hat T_{ab} \; ,
\ee
where \(\hat G_{ab}\) is the Einstein tensor of \(\hat g_{ab}\) and the stress-energy tensor is given by
\be
    \hat T_{ab} = a^2 (\rho + P) \nabla_a \eta \nabla_b \eta + P \;  \hat g_{ab} \; .
\ee
Here \(P\) is the \emph{pressure} and \(\rho\) is the \emph{density} of a perfect fluid, which are related through \(P = w\rho\), and \(w\) is a constant \emph{equation of state} parameter. The \emph{deceleration parameter} is defined by\footnote{Note that the \(q\) used in \cite{KR} is the inverse of the standard convention for the deceleration parameter.}
\be\label{eq:q-defn}
    q \defn 1 - \frac{\ddot a a }{(\dot a)^2} = \frac{1+3w}{2} \; ,
\ee
where the ``overdots'' indicate derivative with respect to the conformal time \(\eta\). The parameter $s$ in \cref{eq:FLRW} is related to the equation of state parameter $w$ and the deceleration parameter \(q\) through
\begin{equation}\label{eq:s-w-q}
s= \frac{2}{3(1+w)} = \frac{1}{1+q} \; .
\end{equation}

 When \(w\) satisfies $ -1/3 < w < \infty$, the deceleration parameter \(q\) is positive and \(0 < s < 1\).  These spacetimes represent an expanding universe whose expansion decelerates towards the future. Examples of such spacetimes include radiation- and dust-filled cosmological solutions for which $w=1/3$ and $w=0$, respectively, and a universe with a stiff-fluid for which $w=1$. Such spacetimes have a null conformal boundary as we will review below (these correspond the case labelled ``F1'' in \cite{Harada:2018ikn}). Spacetimes whose expansions accelerate (\(q < 0\)) in the future evolution, e.g. de Sitter spacetimes, do not posses a null boundary, but instead have a spacelike boundary \cite{Anninos:2010zf,ABK,Harada:2018ikn}. The case \(q=0\) case also has null conformal boundary but scale factor grows exponentially in the conformal time \(\eta\) (see the case labelled ``F2'' in \cite{Harada:2018ikn}). In this paper we will only consider decelerating FLRW spacetimes; note that when \(s = 0\), \cref{eq:FLRW} is simply the Minkowski metric, which we can also include in our analysis.\\

Next, we construct a conformal completion of the decelerating FLRW spacetimes (see Appendix~H of \cite{Carroll-book}). Note that FLRW spacetimes are conformally isometric to a region of Minkowski spacetime. Hence to obtain the conformal completion for FLRW we can follow the same procedure as for Minkowski spacetime. Note that this \emph{does not} imply that FLRW spacetimes are asymptotically flat. To see this, let us consider the conformal completion in more detail.

\begin{figure}[h!]
	\centering
	\includegraphics[width=0.65\textwidth]{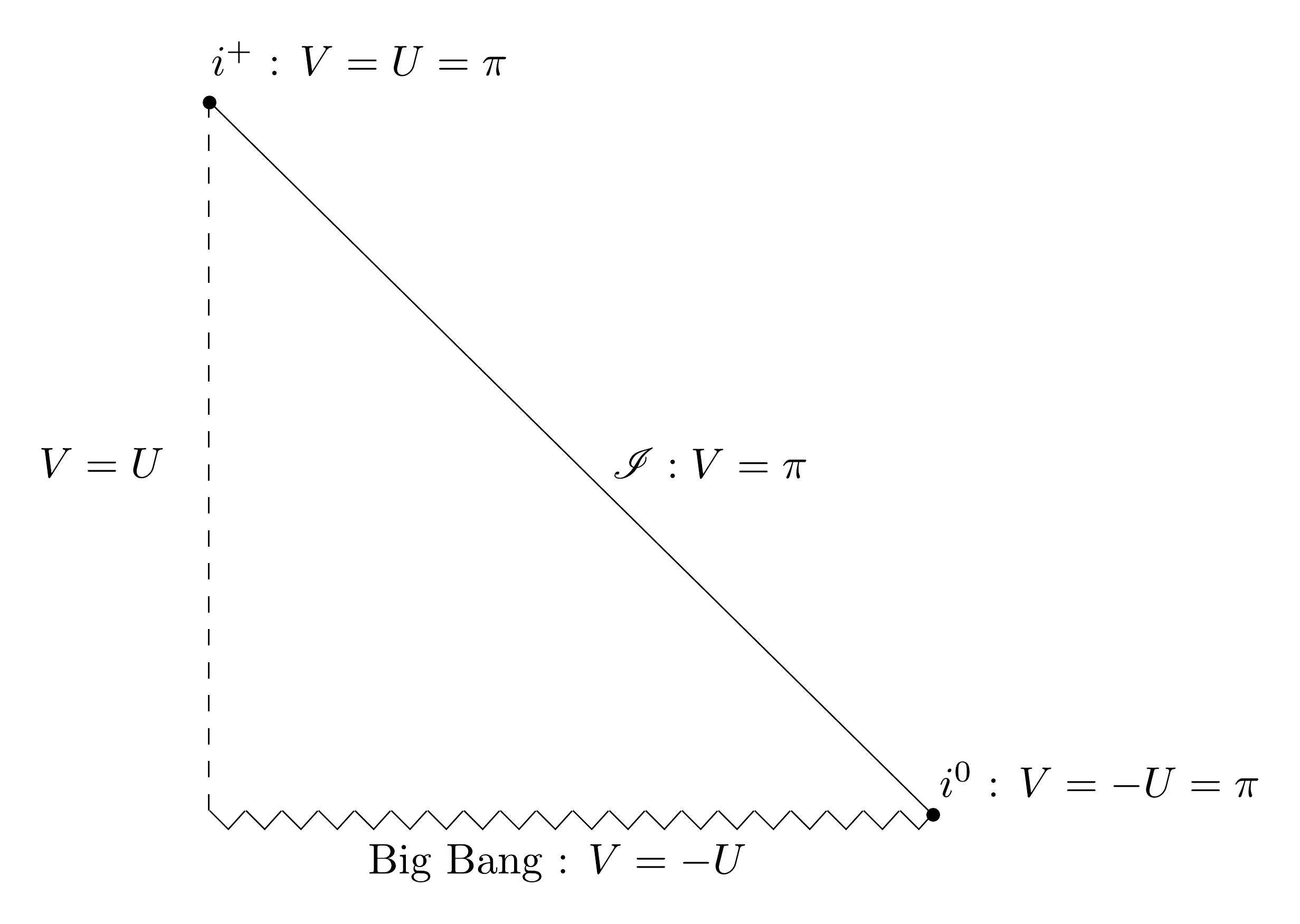}
	\caption{Carter-Penrose diagram for decelerating FLRW spacetimes. The curve \(V=U\) (denoted by a dashed line) is the axis of rotational symmetry and the curve $V=-U$ (denoted by a jagged line) is the Big Bang singularity.}\label{fig:conformaldiagram}
\end{figure}

Choose new coordinates \((T,R)\) and \((U,V)\) in the FLRW spacetime satisfying
\begin{equation}\begin{aligned}
\eta & = \frac{\sin T}{\cos R + \cos T} \eqsp&  r & = \frac{\sin R}{\cos R + \cos T} \\
U & \defn T-R \eqsp& V & \defn T+R
\end{aligned}\end{equation}
and a conformal factor
\be\label{eq:Omega-FLRW}
	\Omega = 2 \left(\cos \tfrac{V}{2} \, \cos \tfrac{U}{2}\right)^{1/(1-s)} \lb( \sin \tfrac{U+V}{2} \rb)^{- s/(1-s)} \; .
\ee
Then, the conformally rescaled metric \(g_{ab} \defn \Omega^2 \hat g_{ab}\) has the line element
\begin{equation}\label{eq:unphys-metric-FLRW}
    ds^2 = - dU dV +  \lb( \sin\tfrac{V-U}{2} \rb)^2 \, S_{AB} dx^A dx^B \, ,
\end{equation}
where we have set \(\eta_0 = 1/2\) for notational convenience in this section. The ranges of the new coordinates inherited from those of \(\eta\) and \(r\) are
\be\label{eq:ranges}\begin{aligned}
	0 < T < \pi \eqsp 0 \leq R < \pi - T \\
	-\pi < U < \pi \eqsp \abs{U} < V < \pi \; .
\end{aligned}\ee
Note that \cref{eq:unphys-metric-FLRW} is the metric of the Einstein static universe which can be extended smoothly to the boundaries of the new coordinates. So the conformal completion of FLRW spacetimes is another spacetime \((M, g_{ab})\) where the manifold \(M\) is a region of the Einstein static universe with boundaries at \(V = -U\) and at \(V = \pi\) and the metric \(g_{ab}\) (\cref{eq:unphys-metric-FLRW}) is smooth everywhere including at the boundaries. From \cref{eq:Omega-FLRW}, we see that \(\Omega\) diverges at the boundary surface \(V= - U\) (corresponding to the Big Bang singularity), and \(\Omega\) vanishes at \(V = \pi\) (corresponding to \emph{null infinity} \(\scri\)). The points \(V = U = \pi\) and \(V = -U = \pi\) represent \emph{future timelike infinity} \(i^+\) and \emph{spatial infinity} \(i^0\), respectively. The Carter-Penrose diagram for the conformally completed spacetime is depicted in \cref{fig:conformaldiagram}.

Despite its simplicity, the above construction can be misleading --- the behaviour of the conformal factor and the stress-energy at \(\scri\) in FLRW spacetimes is crucially different from that in asymptotically flat spacetimes. This difference arises because the FLRW scale factor (which conformally relates the FLRW metric to the Minkowski one) is divergent near \(\scri\) and behaves as
\be\label{eq:scale-asymp}
	a(\eta) = \Omega^{-s} A^s \qquad  \text{ with } \qquad  A \equiv 2 \sin \tfrac{U+V}{2} \hateq 2 \cos \tfrac{U}{2}
\ee
and \(A\) smooth at \(\scri\).

Consider, first, the conformal factor \cref{eq:Omega-FLRW}, which near \(\scri\) (where \(V = \pi\)) behaves as
\be
	\Omega \sim \cos\tfrac{U}{2} \lb( \pi-V \rb)^{1/(1-s)} \eqsp \nabla_a \Omega \sim  \cos\tfrac{U}{2} (\pi - V)^{s/(1-s)} \nabla_a V \; .
\ee
Thus, given that $0\leq s<1$, $\Omega$ is not smooth and \(\nabla_a \Omega \hateq 0\) at \(\scri\) unless \(s=0\). It is tempting to conclude that this is simply a bad choice for $\Omega$ and one should choose another conformal factor which is smooth at \(\scri\). It is easy to check that any new choice of conformal factor \(\Omega' = \omega \Omega\) which is smooth at \(\scri\) and \(\nabla_a \Omega' \not\hateq 0\) requires \(\omega \sim (\pi - V)^{-s/(1-s)}\). But then, the new conformally rescaled metric \(g'_{ab} = \omega^2 g_{ab} \sim (\pi-V)^{-2s/(1-s)} g_{ab}\) diverges at \(\scri\) and one would have to be extremely careful using the tools of differentiable geometry on $\scri$ --- in fact, we would not even be able to conclude that \(\scri\) is a null surface. Thus, we will work with conformal completions in which \(g_{ab}\) is smooth at \(\scri\) and allow the conformal factor \(\Omega\) to not be smooth.

This lack of smoothness of the conformal factor is not a serious drawback. Note that the function \(\Omega^{1-s}\) is smooth and \(\Omega^{1-s} \hateq 0\). Similarly, consider the covector defined by
\be\label{eq:nd}
    n_a \defn \Omega^{-s} \grad_a \Omega = \tfrac{1}{1-s} \nabla_a \Omega^{1-s} \hateq - \tfrac{2^{-s}}{1-s} \lb( \cos \tfrac{U}{2}\rb)^{1-s} \grad_a V  \not\hateq 0 \; .
\ee 
Thus, \(n_a\) is smooth and defines a non-vanishing normal to \(\scri\). In addition, one can verify that $\scri$ is a null surface since $n^a n_a = \mathcal{O}(\Omega^{1-s})$. We emphasize that \(\Omega^{1-s}\) is \emph{not} a new choice of conformal factor, the unphysical metric is still \(g_{ab} = \Omega^2 \hat g_{ab}\) with the line element given by \cref{eq:unphys-metric-FLRW}. Similarly, the pullback of \(g_{ab}\) to \(\scri\) (where \(V = \pi\)) induces a degenerate, smooth metric \(q_{ab}\) with \(q_{ab}n^b \hateq 0\). On cross-sections of \(U = \text{constant}\), this is the metric given by \((\cos \tfrac{U}{2})^2 S_{AB}\).

A direct computation shows that the divergence of the normal \cref{eq:nd} on \(\scri\) is given by
\be
    \grad^a n_a \hateq - 2^{1-s}~ \tfrac{2-s}{1-s} \sin \tfrac{U}{2}\; \lb( \cos \tfrac{U}{2} \rb)^{-s} \; .
\ee
Just as in the asymptotically flat case, we can exploit the freedom in the conformal factor to choose a new normal which is divergence-free; we will call this the \emph{divergence-free conformal frame}. To do this let $\Omega'=\sec \tfrac{U}{2} \Omega$, so that the new normal satisfies
\be\label{eq:nd-div-free}
    n'_a \defn \Omega'^{-s} \grad_a \Omega' \hateq - \tfrac{2^{-s}}{1-s} \grad_a V  \not\hateq 0  \eqsp \grad'_a n'^a \hateq 0 \eqsp n'^a n'_a = \mathcal{O}(\Omega'^{2(1-s)}) \; .
\ee
And further in this choice of conformal factor the new metric is \(g'_{ab} = (\sec \tfrac{U}{2})^2 g_{ab}\) and the induced metric on \(\scri\) is simply the unit round metric.

Another difference with asymptotically flat spacetimes is the presence of matter. For asymptotically flat spacetimes the stress-energy tensor is required to decay as one approaches $\scri$ (specifically, $\Omega^{-2} \hat{T}_{ab}$ should have a limit to $\scri$).\footnote{In asymptotically flat spacetimes, this falloff of the stress-energy holds for conformally invariant fields, e.g., Maxwell fields in electromagnetism. For radiative scalar field solutions \(\hat T_{ab}\) does not decay but has a finite non-zero limit to \(\scri\), see \cite{Flanagan:2015pxa}. As we argue for FLRW spacetimes the behaviour of the stress-energy is even worse as \(\hat T_{ab}\) diverges in the limit to \(\scri\).} This clearly cannot be the case for FLRW spacetimes which have a homogeneous stress-energy. For instance, the trace of the stress-energy tensor is non-vanishing on $\scri$:
\be
	\lim_{\to \scri} 8\pi g^{ab} \hat T_{ab} = \frac{6s(1-2s)}{(1-s)^2} \lb( \sec  \tfrac{U}{2} \rb)^2 \qquad \text{and} \qquad \lim_{\to \scri} 8\pi g'^{ab} \hat T_{ab} = \frac{6s(1-2s)}{(1-s)^2} \; ,
\ee
where the two expressions are evaluated in the initial choice of conformal frame versus the divergence-free frame described above. The situation is worse, however, as certain components of the stress-energy tensor diverge on $\scri$! In particular we find that 
\be\label{eq:T-exp}
    8\pi \hat{T}_{ab} = 2 s \Omega^{2(s-1)} n_a n_b + 2 s \Omega^{s-1} \tau_{(a} n_{b)} + \mc O(1) \, ,
\ee
where \(\mc O(1)\) indicates terms which have a finite limit to \(\scri\). The leading-order divergent term (first term on the right-hand-side above) is independent of the choice of conformal frame, while \(\tau_a\) does depend on the choice of conformal factor (see \cref{eq:tau-transform}). In particular, $\tau_a$ in the initial frame versus the divergence-free frame is given by
\be
    \tau_a \hateq \tan \tfrac{U}{2} \left[\grad_a U + \grad_a V\right] \eqsp \tau'_a \hateq \tan \tfrac{U}{2}  \grad_a V \hateq - 2^{s} (1-s) \tan \tfrac{U}{2} n'_a \;  .
\label{eq:tauFLRW}
\ee\\

In summary, we see that even though FLRW spacetimes are conformal to Minkowski spacetime, their conformal completions have radically different properties at null infinity. Motivated by the properties of \(\scri\) in exact FLRW spacetimes discussed above, we now define a class of spacetimes whose behaviour at null infinity is similar to those of the FLRW spacetimes.

\section{Spacetimes with a cosmological null asymptote}\label{sec:dna}

In this section we define a class of spacetimes with a \emph{cosmological null asymptote} and discuss its geometric structure at null infinity in detail. The spacetimes in this class behave like a decelerating FLRW spacetime at null infinity, similar to how asymptotically flat spacetimes behave like Minkowski spacetime at null infinity. Similarities and differences with asymptotically flat spacetimes are highlighted along the way.

Finally, given that decelerating FLRW spacetimes are the only explicit example satisfying this definition, we construct a large class of spacetimes that have a cosmological null asymptote. In fact, this construction illustrates that the class of spacetimes with a cosmological null asymptote is even larger than the class of asymptotically flat spacetimes.

\begin{definition}[Cosmological null asymptote]\label{def:dna}
A (physical) spacetime \((\hat M, \hat g_{ab})\) satisfying the Einstein equation $\hat{G}_{ab} = 8 \pi \hat{T}_{ab}$ with stress-energy tensor \(\hat T_{ab}\) admits a \emph{cosmological null asymptote} at infinity if there exists another (unphysical) spacetime $(M, g_{ab})$ with boundary $\scri \cong \bb R \times \bb S^2$ and an embedding of $\hat{M}$ into $M - \scri$ such that\footnote{We use the convention whereby \(\hat M\) is identified with its image in \(M\) under the embedding.}
\begin{condlist}
\item There exists a function $\Omega > 0$ which is smooth on \(\hat M\) and can be, at least, \emph{continuously} extended to \(\scri\) such that
	\begin{condlist}
	\item \(\Omega \hateq 0\) and $g_{ab}= \Omega^2 \hat{g}_{ab}$ is smooth and nondegenerate on \(M\), where \(\hateq\) stands for ``equals when evaluated on \(\scri\)", and, \label{di:conformal-relation}

	\item for some constant \(0 \leq s < 1\), \(\Omega^{1-s}\) is smooth on \(M\), and $n_a \defn \tfrac{1}{1-s} \grad_a \Omega^{1-s}$ is nowhere vanishing on $\scri$. \label{di:smooth-things}
	\end{condlist}

\item The stress-energy tensor \(\hat T_{ab}\) is such that \label{di:T-cond}
	\begin{condlist}
	\item \(\lim\limits_{\to \scri} g^{ab}\hat T_{ab}\) exists, and, \label{di:lim-trace-T}
	\item \( \lim\limits_{\to \scri} \Omega^{1-s} \lb[ 8\pi \hat T_{ab} - 2s \Omega^{2(s-1)} n_a n_b  \rb] \hateq 2 s \tau_{(a} n_{b)} \), for some smooth \(\tau_a\) on \(\scri\) . \label{di:lim-T}
	\end{condlist}
\end{condlist}
\end{definition}

The \cref{di:conformal-relation} has the usual meaning that the boundary \(\scri\) is ``infinitely far" from all points in the physical spacetime \(\hat M\). That the boundary \(\scri\) is null, and not spacelike or timelike, follows from other conditions in \cref{def:dna} as we will show later in this section. \Cref{di:smooth-things} ensures that there is some (fractional) power of the conformal factor \(\Omega\) which is smooth at \(\scri\) with non-vanishing gradient which can be used as a normal to \(\scri\). Since \(\Omega^{1-s}\) is assumed to be smooth on \(M\), the conformal factor \(\Omega\) is smooth when \(\frac{1}{1-s}\) is an integer, otherwise it is only differentiable \(\floor{\frac{1}{1-s}}\)-times at \(\scri\). For the case of interest where, \(0 \leq s < 1\), we see that \(\Omega\) is always at least once-differentiable and \(\grad_a \Omega = \Omega^s n_a \hateq 0\).

The \cref{di:T-cond} ensures that the stress-energy tensor is ``suitably regular" at \(\scri\) as suggested by the behaviour of the stress-energy for decelerating FLRW spacetimes detailed in \cref{sec:conformalcompletionFLRW}. In particular, \cref{di:lim-trace-T} will be essential to showing that the boundary \(\scri\) is a null hypersurface in \(M\) (see \cref{eq:nisnull} below). \Cref{di:lim-T} places restrictions on the singular behaviour in the stress-energy tensor at \(\scri\) --- the leading-order singular term given by \(2s \Omega^{2(s-1)} n_a n_b\) is ``universal" while the next-order singular term is encoded in \(\tau_a\). This will be crucial to show that \(\scri\) is, in fact, geodesic and has vanishing shear and expansion (see \cref{eq:gradneqn} below).

Comparing to the special case of decelerating FLRW spacetimes, we may interpret the parameter \(s\) as encoding the ``asymptotic equation of state" or ``asymptotic deceleration" (\cref{eq:s-w-q}) and \(\Omega^{-s}\) as the ``asymptotic scale factor" (\cref{eq:scale-asymp}) up to functions that are smooth at \(\scri\). Note that for \(s=0\), we have \(n_a = \grad_a \Omega\) just as in the asymptotically flat case. However, \cref{def:dna} allows for a \(\hat T_{ab}\) which has a limit to \(\scri\) while standard definitions of asymptotic flatness require instead the stronger condition that \(\lim\limits_{\to \scri} \Omega^{-2} \hat T_{ab}\) exists. For the results of this section this stronger falloff condition will play no essential role and one is free to think of the \(s=0\) case as the asymptotically flat case.

\begin{remark}[Completeness of the asymptote \(\scri\)]\label{rem:completeness}
The standard definition of asymptotic flatness also assumes that the integral curves of \(n^a\), in a suitable choice of conformal factor \(\Omega\), are complete (see \S~11.1 of \cite{Wald-book}). That is, null infinity of asymptotically flat spacetimes is ``as big as" null infinity of Minkowski spacetime. This global condition is essential, for instance, in the definition of black hole spacetimes and for describing the asymptotic symmetry group, see \cite{GH,Foster} (see also \S~IV.C. of \cite{ABK} for the role of the global topology of the asymptotic boundary in the asymptotically de Sitter case). We will not be concerned with these global issues here and we leave the question of completeness of \(\scri\) unaddressed in \cref{def:dna}. In this sense the \(\scri\) in \cref{def:dna} is only a ``local asymptote" and not necessarily a ``global asymptotic boundary".  
\end{remark}

We now turn to showing that the asymptote \(\scri\) is indeed null as a consequence of the Einstein equation and the conditions in \cref{def:dna}. Using the conformal relation \cref{di:conformal-relation} between $g_{ab}$ and $\hat{g}_{ab}$, we can write the Einstein equation as:
\be\label{eq:EE}\begin{split}
	8\pi \hat{T}_{ab} & = G_{ab} + 2 \Omega^{-1} \left(\grad_a \grad_b \Omega - g_{ab} \grad^c \grad_c \Omega \right) + 3 \Omega^{-2} g_{ab} \grad^c \Omega \grad_c \Omega \\
	& = G_{ab} + 2 \Omega^{s-1} \left(\grad_a n_b  - g_{ab} \grad^c n_c \right) + \Omega^{2(s-1)} \left(2 s n_a n_b + \left(3-2s\right) g_{ab} n^c n_c \right) \; .
\end{split}\ee
Since \(\Omega\) need not be twice-differentiable at \(\scri\) for a general \(s\), in the second line we have rewritten this in terms of \(n_a\) and \(\Omega^{1-s}\) which are both smooth on \(\scri\) (\cref{di:smooth-things}). Contracting \cref{eq:EE} with $g^{ab}$ and multiplying by $\Omega^{2(1-s)}$ we find
\be\label{eq:nisnull}
	n^a n_a = \tfrac{1}{2-s} \lb [\Omega^{1-s} \grad^a n_a + \Omega^{2(1-s)} \lb( \tfrac{4\pi}{3} g^{ab}\hat{T}_{ab} + \tfrac{1}{6} R \rb) \rb] \hateq 0 \; ,
\ee
where the final equality on \(\scri\) uses \cref{di:lim-trace-T}. Thus, $n^a$ is null on $\scri$ and $\scri$ is a null hypersurface in \(M\). Consequently, the null normal $n^a$ is also tangential to $\scri$ and the pullback of \(g_{ab}\) defines an intrinsic metric $q_{ab} \defn \pb{g_{ab}}$ on $\scri$ satisfying \(q_{ab} n^b \hateq 0\), i.e. \(q_{ab}\) is degenerate with signature $(0, +, +)$. \\

There is considerable freedom in the choice of the conformal factor $\Omega$ relating the physical and unphysical spacetimes. Given a choice of $\Omega$ satisfying the conditions in \cref{def:dna}, let $\Omega'=\omega \Omega$ be another conformal factor allowed by \cref{def:dna}. Since \(g'_{ab} = \Omega'^2 \hat g_{ab} = \omega^2 g_{ab}\), \cref{di:conformal-relation} implies that $\omega > 0$ is smooth on $M$. Note that any power of \(\omega\) is also smooth on \(M\) since \(\omega\) does not vanish. Further, we have
\be\label{eq:nd-transform}
	n'_a = \omega^{1-s} n_a + \Omega^{1-s} \omega^{-s} \nabla_a \omega 
\ee 
and thus \(n'_a\) satisfies \cref{di:smooth-things}. Similarly, \cref{di:lim-trace-T,di:lim-T} are satisfied in the new completion with
\be\label{eq:tau-transform}
	\tau'_a \hateq \tau_a - 2 \grad_a \ln \omega \; .
\ee
Thus, the freedom in the choice of conformal factor is any smooth function \(\omega > 0\) on \(M\).

We can use this freedom to choose a conformal factor so that the null normal $n'^a = g'^{ab}n'_b $ is divergence-free on $\scri$. We proceed as follows
\begin{subequations}\begin{align}
	n'^a & = \omega^{-1-s} n^a + \Omega^{1-s} \omega^{-2-s} g^{ab} \grad_b \omega \label{eq:nU-transform} \\
	\implies \grad'_a n'^a & \hateq \left(4-2 s \right) \omega^{-2-s} \Lie_n \omega + \omega^{-1-s} \grad_a n^a \label{eq:div-n-transform} \; .
\end{align}\end{subequations}
We choose $\omega$ so that it solves
\be\label{eq:constrainomega}
	\Lie_n \ln\omega \hateq - \tfrac{1}{2(2-s)} \grad_a n^a \, .
\ee
Since \cref{eq:constrainomega} is an ordinary differential equation along each integral curve of $n^a$ it always has solutions locally. Since, we do not impose any ``global" conditions on the asymptote \(\scri\) (see \cref{rem:completeness}) the existence of such local solutions suffices for our purposes.
Then, from \cref{eq:div-n-transform} it follows that \(\nabla'_a n'^a \hateq 0\), i.e. the new null normal \(n'^a\) is divergence-free on \(\scri\). Further, from \cref{eq:nisnull} we see that the divergence-free normal is null in a first-order neighbourhood of \(\scri\), i.e. \(\lim\limits_{\to \scri} \Omega'^{s-1} n'_a n'^a \hateq 0\). Henceforth, we will assume that some such \emph{divergence-free conformal frame} has been chosen and drop the ``prime" from the notation.\\

Next, multiplying \cref{eq:EE} by $\Omega^{1-s}$ and using \cref{di:lim-T} we obtain:  
\begin{align}\label{eq:gradneqn}
	\grad_a n_b & = \tfrac{1}{2} \Omega^{1-s} \lb[ 8\pi \hat{T}_{ab} - 2 s \Omega^{2(s-1)} n_a n_b \rb] - \tfrac{1}{2} \Omega^{1-s} G_{ab}  -\tfrac{1}{2} g_{ab} \lb[ \Omega^{s-1} (3-2s) n^c n_c - 2 \grad^c n_c \rb]  \notag \\
	& \hateq s \tau_{(a} n_{b)} \; .
\end{align}
Further, contracting \cref{eq:gradneqn} with \(g^{ab}\) (or alternatively, contracting \cref{di:lim-T} with \(g^{ab}\) and using \cref{di:lim-trace-T}) we have (in any divergence-free conformal frame)
\be\label{eq:n-tau}
	n^a \tau_a \hateq 0 \; .
\ee
Note that, unlike the case of asymptotically flat spacetimes, \(\grad_a n_b \not\hateq 0\), i.e. the Bondi condition is not satisfied when \(s \neq 0\) due to the matter term \(\tau_a\). However, the pullback of \(\grad_a n_b\) to \(\scri\) still vanishes. Thus, \(\scri\) has vanishing shear and expansion, is generated by affinely-parametrised null geodesics with tangent \(n^a\), and the degenerate metric \(q_{ab}\) on \(\scri\) satisfies
\be\label{eq:lie-n-q}
	\Lie_n q_{ab} \hateq 0 \; ,
\ee
that is, \(q_{ab}\) is the lift to \(\scri\) of a positive-definite metric on the space of generators \(S \cong \bb S^2\).

Even in a divergence-free conformal frame, there is a residual conformal freedom $\Omega \mapsto \Omega'= \omega \Omega$ with $\Lie_n \omega \hateq 0$. Note that, from \cref{eq:nU-transform,eq:tau-transform}, we see that \cref{eq:n-tau} holds for any such choice of \(\omega\). This conformal freedom can be used to restrict the metric \(q_{ab}\) as follows. Let \(S \cong \bb S^2\) be some cross-section of \(\scri\). It follows from the \emph{uniformization theorem} (for instance see Ch.~8 of \cite{Bieri-CK-ext0})\footnote{The uniformization theorem is a \emph{global} result depending on the topology of the \(2\)-dimensional space. Locally, all metrics of a particular signature on a \(2\)-surface are conformally-equivalent, Problem~2 Ch.~3 of \cite{Wald-book}.} that any metric on \(S\) is conformal to the unit round metric on \(\bb S^2\) (that is, the metric with constant Ricci scalar with value \(2\)). Thus, we can always choose the conformal factor so that the metric \(q_{ab}\) on this cross-section \(S\) is also the unit round metric of \(\bb S^2\). From \cref{eq:lie-n-q}, it then follows that this holds on any cross-section. This defines a \emph{Bondi conformal frame}. Thus on \(\scri\), in the Bondi conformal frame the divergence of the normal \(n^a\) vanishes and the induced metric \(q_{ab}\) is that of a unit round \(2\)-sphere.\\

In summary, we see that null infinity for spacetimes satisfying \cref{def:dna} has many similarities to those of asymptotically flat spacetimes:
\begin{enumerate*}[label=(\roman*)]
    \item the asymptote \(\scri\) is indeed null, as a consequence of the Einstein equation and the conditions on the stress-energy in \cref{def:dna};
    \item One can pick the conformal factor so that the normal \(n^a\) is divergence-free and the induced metric \(q_{ab}\) coincides with that of a unit round \(2\)-sphere.
\end{enumerate*}
However, there are also some very crucial differences:
\begin{enumerate*}[label=(\alph*)]
    \item The stress-energy tensor \(\hat T_{ab}\) diverges at \(\scri\), as specified in \cref{di:lim-T}, similar to its behaviour in exact FLRW spacetimes as detailed in \cref{sec:conformalcompletionFLRW};
    \item the conformal factor \(\Omega\) is not smooth at \(\scri\) --- its degree of smoothness is parametrized by a fractional parameter \(s\) so that \(\Omega^{1-s}\) is smooth and \(\tfrac{1}{1-s}\nabla_a \Omega^{1-s}\) defines the non-vanishing normal at \(\scri\);
  \item Due to the diverging stress-energy tensor, the Bondi condition cannot be satisfied at \(\scri\) (see \cref{eq:gradneqn}).
\end{enumerate*}

As we shall show in \cref{sec:symm} below, the above similarities and differences manifest themselves directly in the asymptotic symmetry algebra of spacetimes with a cosmological null asymptote --- the symmetry algebra has a structure quite similar to the usual BMS algebra of asymptotically flat spacetimes but is \emph{not isomorphic} to the BMS algebra when \(s\neq 0\).

\subsection{Existence of a large class of spacetimes with a cosmological null asymptote}\label{sec:manyspacetimes}

In the asymptotically flat case, the existence of a large class of spacetimes (beyond the known exact solutions like Kerr-Newman or Vaidya spacetimes) satisfying the requisite properties is supported by many nontrivial results. Geroch and Xanthopoulos showed that in vacuum general relativity asymptotic flatness is linearisation stable, i.e., linear perturbations with initial data of compact support on some Cauchy surface preserve the conditions for asymptotic flatness at null infinity \cite{GX} (see also \cite{HI}). In full nonlinear general relativity the existence of a large class of asymptotically flat spacetimes is supported by many results; see \cite{CK,CD,Bieri-CK-ext-thesis,Bieri-CK-ext0,Zipser,Bieri-CK-ext1}.

However, the only exact solutions (that we are aware of) satisfying \cref{def:dna}, for a given value of \(s\), are the decelerating FLRW spacetimes detailed in \cref{sec:conformalcompletionFLRW}. Thus, in this section we turn our attention to showing that there does indeed exist a large class of spacetimes satisfying \cref{def:dna}. Generalizing the stability results mentioned above promises to be a difficult task. For the case of FLRW with a positive cosmological constant some global stability results have been proven in \cite{Rodnianski:2009de,Speck:2011jr}, as well as for some non-accelerating cases but with spatial topology of a \(3\)-torus \cite{Oliynyk:2020gcg,Fajman:2020tcf}. Global stability has also been shown for the Einstein-Vlasov equations for initial data of compact support enjoying certain symmetries \cite{DR}. However, no global stability results exist for the decelerating FLRW spacetimes of interest here (see \cref{sec:conformalcompletionFLRW}). Even the linearization stability result of Geroch and Xanthopoulos does not seem to generalize straightforwardly. The presence of matter complicates the linearized equations of motion considerably. Additionally, one would also need to suitably modify the gauge choice used in \cite{GX}.\footnote{Note that the frequently used harmonic gauge is not a suitable choice at null infinity even in asymptotically flat spacetimes in \(4\)-dimensions \cite{HI}: in coordinate form, the solutions in harmonic gauge behave like $\ln r$ in the Bondi coordinate \(r\) near \(\scri\). The transverse-traceless gauge in FLRW is also not suitable since it is adapted to the homogenous spatial slices which do not reach null infinity.}  Instead we will show that given any asymptotically flat spacetime, possibly with some matter stress-energy tensor, we can construct a different physical spacetime with a different stress-energy tensor satisfying \cref{def:dna}. The reader not interested in the details of this construction may safely skip this subsection.\\

Let \((\af M, \af g_{ab})\) be an asymptotically flat spacetime with stress-energy tensor \(\af T_{ab}\), and let \(\af \Omega\) be a smooth conformal factor for its completion to the unphysical spacetime \((M, g_{ab})\) with null infinity \(\scri\), then we have
\be\label{eq:asymp-flat-condn}
	g_{ab} = \af \Omega^2 \af g_{ab} \eqsp \af \Omega \hateq 0 \eqsp \af n_a = \nabla_a \af \Omega \not\hateq 0 \eqsp \text{and} \quad \lim\limits_{\to\scri} \af \Omega^{-2} \af T_{ab} \text{ exists} \;.
\ee
Further, let us assume that \(\af \Omega\) is chosen such that \(\nabla_a \af n_b \hateq 0\) and \(q_{ab} \hateq \pb{g_{ab}}\) is the unit round metric.

For \(0\leq s < 1\), we want to construct a physical spacetime \((\hat M, \hat g_{ab})\) with stress-energy \(\hat T_{ab}\) and conformal factor \(\Omega\) so that it has the same conformal completion \((M, g_{ab})\) in a neighbourhood of \(\scri\) such that (at least some part of) \(\scri\) is a cosmological null asymptote for \((\hat M, \hat g_{ab})\) in the sense of \cref{def:dna}. We proceed as follows. Let \(A > 0\) be some smooth function on \(M\) which satisfies \(\af n^a \grad_a A \hateq 0\) and define \(\Omega\) through the relation
\be
	\af \Omega = \Omega^{1-s} A^s \; .
\ee
Then, since \(s < 1\), we have \(\Omega \hateq 0\) and  \(n_a \defn \Omega^{-s}\nabla_a \Omega\) satisfies
\be\label{eq:new-n-condn}
	n_a \hateq \tfrac{1}{1-s} A^{-s} \af n_a \not\hateq 0 \eqsp \lim\limits_{\to\scri} \Omega^{s-1} n^a n_a \hateq \nabla_a n^a \hateq 0 \eqsp \nabla_a n_b \hateq -2s n_{(a} \grad_{b)} \ln A \; .
\ee
Next, in a neighbourhood of \(\scri\) we define the ``asymptotic scale factor" \(\alpha\) by
\be
	\alpha \defn \Omega^{-s} A^s  \; .
\ee
Note that \(\alpha\) is not smooth at \(\scri\), just as the scale factor of FLRW spacetimes is not. Now we can finally construct the physical metric \(\hat g_{ab}\) by
\be\label{eq:g-from-af}
	\hat g_{ab} \defn \Omega^{-2} g_{ab} = \alpha^2 \af g_{ab}
\ee 
and the stress-energy tensor \(\hat T_{ab}\) by
\be\label{eq:T-from-af}\begin{split}
	8\pi \hat T_{ab} & \defn 8\pi \af T_{ab} - 2 \alpha^{-1} \lb[ \nabla_a \nabla_b \alpha - g_{ab} \nabla^2 \alpha + \Omega^{s-1} (2 \nabla_{(a} \alpha n_{b)} + \Lie_n \alpha g_{ab}) \rb] - 3 \alpha^{-2} g_{ab} \nabla^c \alpha \nabla_c \alpha \\
	& = 8\pi \af T_{ab} - 2 \alpha^{-1} \lb( \hat\nabla_a \hat\nabla_b \alpha - \hat g_{ab} \hat\nabla^2 \alpha \rb) -3 \alpha^{-2} \hat g_{ab} \hat\nabla^c \alpha \hat\nabla_c \alpha \; ,
\end{split}\ee
where in the second line we have converted to the physical covariant derivative \(\hat\grad\). With the above definitions it is straightforward to check  that
\be
	\af G_{ab} = 8\pi \af T_{ab} \implies \hat G_{ab} = 8 \pi \hat T_{ab} \; .
\ee
So far, we have merely performed some conformal transformations starting from some arbitrary asymptotically flat spacetime to construct a new metric \(\hat g_{ab}\) and a new stress-energy tensor $\hat{T}_{ab}$ such that Einstein equation is satisfied. Therefore, at this point, it is not at all obvious that $\hat{T}_{ab}$ satisfies the conditions on the stress-energy tensor in \cref{def:dna}. We will now show that the conditions on $A$ naturally ensure this. We start by rewriting \cref{eq:T-from-af} in terms of the smooth function \(A\):
\be\begin{split}
	8 \pi \hat T_{ab} & = 8\pi \af T_{ab} + 2 s \Big[ \Omega^{2(s-1)} n_a n_b + \Omega^{s-1} \nabla_a n_b - 2 (1 - s) \Omega^{s-1} n_{(a} \nabla_{b)} \ln A \\
	&\quad  - s \nabla_a \ln A \nabla_b \ln A - \nabla_a \nabla_b \ln A \Big] + s g_{ab} \Big[ (4 - 3 s) \Omega^{2 (s-1)} n_{c} n^{c}  - 2 \Omega^{s-1} \nabla_{c}n^{c} \\
	&\quad - 2 (1 - s) \Omega^{s-1} \Lie_n \ln A + 2 \nabla^2 \ln A - s \nabla_c \ln A \nabla^c \ln A \Big] \, .
\end{split}\ee
Since \(\lim\limits_{\to\scri} \af \Omega^{-2} \af T_{ab}\) exists, we find that 
\be\begin{split}
	\lim\limits_{\to\scri} 8 \pi g^{ab} \hat T_{ab} & = 6s \Big[ (3-2s) \Omega^{2(s-1)} n_a n^a - \Omega^{s-1} \nabla_a n^a \\
	&\quad  - 2 (1 - s) \Omega^{s-1} \Lie_n \ln A - s \nabla_a \ln A \nabla^a \ln A +  \nabla^2 \ln A \Big]  \, .
\end{split}\ee
Using \cref{eq:new-n-condn}, it is clear that this limit exists so that \cref{di:lim-trace-T} is satisfied.
Furthermore, we also obtain
\be
	\lim\limits_{\to\scri} \Omega^{1-s} \lb[ 8 \pi \hat T_{ab} - 2s \Omega^{2(s-1)}n_a n_b \rb] = -4 s n_{(a} \grad_{b)} \ln A  \; ,
\ee
so that
\be
	\tau_a \hateq -2 \nabla_a \ln A \eqsp n^a \tau_a \hateq 0 \; .
\ee
As a result, \cref{di:lim-T} is satisfied and this illustrates nicely the universal behaviour of the leading order divergent part of the stress-energy tensor given by $ 2s \Omega^{2(s-1)}n_a n_b$. As anticipated, the subleading divergent piece depends on $A$ and is therefore not universal. 

A few comments are in order. 
First of all, spacetimes constructed following the above procedure always satisfy \(\grad_{[a} \tau_{b]} \hateq 0\) so this construction does not generate all spacetimes allowed by \cref{def:dna}. Thus the class of spacetimes satisfying \cref{def:dna} is at least as big as the class of asymptotically flat spacetimes.
Second, depending on the specifics of the function $A$, the \(\hat T_{ab}\) constructed in this manner may or may not satisfy any (desirable) energy conditions. 
Third, given that FLRW spacetimes are conformally flat, they obviously can also be constructed using this procedure; the explicit construction is detailed in \cref{sec:FLRW-BS} in the Bondi-Sachs coordinates of the physical spacetime.

\section{Universal structure on \(\scri\) and the asymptotic symmetry algebra}
\label{sec:symm}

In this section, we describe the universal structure emerging from \cref{def:dna}. Next, we derive the asymptotic symmetry algebra, which is the collection of all infinitesimal diffeomorphisms of $\scri$ that preserve the universal structure. This algebra shares similarities with the BMS algebra of asymptotically flat spacetimes, but is not isomorphic to it. In particular, the asymptotic symmetry algebra of spacetimes with a cosmological null asymptote does not contain a translation subalgebra. \\

Given a spacetime satisfying \cref{def:dna}, the asymptote \(\scri\) is null with (local) topology \(\bb R \times \bb S^2\) and comes equipped with the intrinsic fields \((q_{ab}, n^a)\) as discussed above. Due to the freedom in the choice of the conformal factor $\Omega \mapsto \omega \Omega$, these fields are determined by the physical spacetime only up to equivalence under the conformal transformations \((q_{ab}, n^a) \mapsto (\omega^2 q_{ab}, \omega^{-1-s}n^a)\).

While it may seem that different spacetimes give rise to different equivalence classes of \((q_{ab}, n^a)\) on \(\scri\), these fields are in fact universal, i.e, independent of the chosen physical spacetime. To see this, consider two physical spacetimes satisfying \cref{def:dna} (for a given value of \(s\)) with conformal completions with conformal factors \(\Omega\) and \(\Omega'\) and their corresponding null infinities \(\scri\) and \(\scri'\), respectively. In both spacetimes we choose the conformal factors so that we are in the Bondi conformal frame, i.e., the normals \(n^a\) and \(n'^a\) are divergence-free and the metrics \(q_{ab}\) and \(q'_{ab}\) are the unit round \(2\)-sphere metrics. Then, there clearly exists a (not unique) diffeomorphism between the space of generators \(S\) and \(S'\) of \(\scri\) and \(\scri'\), respectively, such that the induced metric \(\underline{q}_{ab}\) on \(S\) maps to the induced metric \(\underline{q}'_{ab}\) on \(S'\) under this diffeomorphism. Then, using this diffeomorphism we can identify the null generators of \(\scri\) and \(\scri'\). Now let \(u\) and \(u'\) be parameters along the null generators of \(\scri\) and \(\scri'\), respectively, such that \(n^a \nabla_a u \hateq 1\) and \(n'^a \nabla_a u' \hateq 1\). We can identify the points along the null generators by the diffeomorphism \(u \hateq u'\). Since \(\scri\) and \(\scri'\) are both topologically \(\bb R \times \bb S^2\), we have setup a diffeomorphism between \(\scri\) and \(\scri'\) which identifies their normals and induced metrics in a fixed conformal frame. Varying the conformal factor then also identifies their equivalence classes under conformal transformations described above. This identification can be done for any two, and hence \emph{all}, spacetimes in the class defined by \cref{def:dna} for any given value of \(s\).

Thus for a given \(s\), the \emph{universal structure} common to all spacetimes satisfying \cref{def:dna} is given by:
\begin{enumerate}
\item a smooth manifold \(\scri \cong \bb R \times \bb S^2\),
\item an equivalence class of pairs $(q_{ab}, n^a)$ on \(\scri\) where \(n^a\) is a vector field and \(q_{ab}\) is a (degenerate) metric with $q_{ab}n^b \= 0$ and $\Lie_n q_{ab} \=0 $, and,
\item any two members of the equivalence class are related by the map \((q_{ab}, n^a) \mapsto (\omega^2 q_{ab}, \omega^{-1-s}n^a)\) for some $\omega > 0$ satisfying $\Lie_n \omega \= 0$.
\end{enumerate}
This universal structure is common to \emph{all} spacetimes satisfying \cref{def:dna} with a given value of the parameter \(s\). Physically different spacetimes are distinguishable only in the ``next-order'' structure such as the matter field $\tau_a$, the derivative operator on $\scri$ induced by the derivative operator compatible with $g_{ab}$ on $M$ and the curvature tensors.\\

The \emph{asymptotic symmetry algebra} is the algebra of infinitesimal diffeomorphisms of \(\scri\) which preserves this universal structure.
Recall that we focus only on the asymptotic symmetry algebra rather than the asymptotic symmetry group as this requires global conditions (see \cref{rem:completeness}).
For asymptotically flat spacetimes, the asymptotic symmetry algebra is the BMS algebra. As we show below, despite the low differentiability of the conformal factor $\Omega$ and the presence of matter fields on $\scri$, the asymptotic symmetry algebra for spacetimes with a cosmological null asymptote is very similar to the BMS algebra.

Concretely, the asymptotic symmetry algebra consists of all smooth vector fields $\xi^a$ on \(\scri\) that map one pair $(q_{ab}, n^a)$ to another equivalent pair $(q'_{ab}, n'^a)$ within the universal structure.  The conditions on $\xi^a$ that ensure this are
\begin{equation}\label{eq:conditionsxi}
\Lie_\xi q_{ab} \= 2 \alpha_{(\xi)} q_{ab} \qquad \text{and} \qquad \Lie_\xi n^a \= - (1+s) \, \alpha_{(\xi)} n^a  \, ,
\end{equation}
where $\alpha_{(\xi)}$ is any function (depending on the vector field \(\xi^a\)) on $\scri$ such that $\Lie_n \alpha_{(\xi)} \= 0$. The vector fields $\xi^a$ satisfying \cref{eq:conditionsxi} form a Lie algebra which we denote by \(\b_s\).\\ 

Next we explore the structure of this Lie algebra. Consider first vector fields of the form $\xi^a \hateq f n^a$, which satisfy \cref{eq:conditionsxi} if and only if $\Lie_n f \= 0$ and \(\alpha_{(fn)} \hateq 0\). It can be verified that such vector fields form an infinite-dimensional abelian subalgebra \(\s_s \subset \b_s\) of \emph{supertranslations}. Further, since \(\Lie_n f \hateq 0\), \(f\) is a function on the space of generators of \(\scri\) which is topologically \(\bb S^2\). Note that under a change of conformal factor \(n^a \mapsto \omega^{-1-s} n^a\) and as a result a fixed supertranslation \(\xi^a \hateq f n^a\) is specified by a function \(f\) on \(\bb S^2\) which transforms as \(f \mapsto \omega^{1+s} f\). Thus, the supertranslation subalgebra is parametrized by smooth functions on \(\bb S^2\) with conformal weight \(1+s\).

Next, the Lie bracket of any vector field \(\xi^a\) in \(\b_s\) and a supertranslation \(fn^a\) is given by
\begin{equation}\label{eq:Lieideal}
[\xi, fn]^a = \left(\Lie_\xi f - (1+s)  \alpha_{(\xi)} \, f  \right) n^a \, .
\end{equation}
Using \cref{eq:conditionsxi} it can be shown that the right-hand-side of the above is also a supertranslation, thus the supertranslation subalgebra \(\s_s\) is a Lie ideal in \(\b_s\).
Therefore, we can quotient $\b_s$ with $\s_s$ to get a Lie algebra $\b_s / \s_s$. To get a concrete realization of this quotient, note that supertranslations preserve any null generator of \(\scri\), and thus the quotient \(\b_s / \s_s\) can be identified with vector fields on the space $S$ of generators of $\scri$. Further, since $\Lie_\xi n^a \propto n^a$ and \(\alpha_{(fn)} \hateq 0\) on $\scri$, from \cref{eq:conditionsxi} it is clear that any \(X^a \in \b_s / \s_s\) satisfies
\begin{equation}
    \Lie_X \underline{q}_{ab} \= 2 \alpha_{(X)} \underline{q}_{ab}  \, ,
\end{equation}
where $\underline{q}_{ab}$ is the positive-definite metric on the space of generators $S \cong \bb S^2$ whose lift to $\scri$ yields $q_{ab}$. In other words, $X^a$ is a conformal Killing vector field on \(\bb S^2\). Given that 2-spheres carry a unique conformal structure, the Lie algebra $\b_s / \s_s$ is the algebra of conformal isometries of the unit \(2\)-sphere which is isomorphic to the Lorentz algebra \(\mf{so}(1,3)\). 
Hence, for spacetimes with a cosmological null asymptote the asymptotic symmetry algebra is given by the semi-direct sum of supertranslations and the Lorentz algebra, i.e.,
\be
    \b_s \cong \mf{so}(1,3) \ltimes \s_s \, .
\ee\\

As is to be expected, for \(s=0\) the universal structure is identical to that of asymptotically flat spacetimes and the algebra \(\b_{s=0} \cong \mf{bms}\) is the BMS algebra. For \(s\neq 0\) the symmetry algebra is similar to the BMS algebra: it is a semi-direct sum of the Lorentz algebra with an infinite-dimensional abelian algebra of supertranslations parametrized by conformally-weighted functions on \(\bb S^2\). However, the algebras \(\b_s\) for \(s\neq 0\) are \emph{not} isomorphic to the BMS algebra. This is because the supertranslation functions in \(\s_s\) have conformal weight \(1+s\), while the BMS supertranslations have conformal weight \(1\). This difference is ultimately due to the fact that the gradient of the conformal factor \(\nabla_a \Omega\) does not define a ``good'' non-zero normal to \(\scri\) when \(s \neq 0\) and one needs to use \(\tfrac{1}{1-s} \nabla_a \Omega^{1-s}\) as the normal instead.

The difference in the conformal weight of the functions parametrizing the supertranslations makes the structure of asymptotic symmetry algebra \(\b_{s\neq 0}\) very different from that of the BMS algebra. Note that the Lie bracket \cref{eq:Lieideal}, gives an action of the Lorentz algebra on the supertranslation function \(f\) with conformal weight \(1+s\). It can be shown that functions with different conformal weights are different (infinite-dimensional irreducible) representations of the Lorentz algebra (see \cite{HNP}). In particular, the representation corresponding to functions of conformal weight \(1\) has a \(4\)-dimensional Lorentz-invariant space of functions which are the preferred Lie subalgebra of \emph{translations} in the BMS algebra --- when the metric \(q_{ab}\) is a unit round metric these are functions spanned by the first four spherical harmonics. However, in the general case where \(s \neq 0\) there is no finite-dimensional Lorentz-invariant space of functions, and hence no finite-dimensional preferred Lie subalgebra of translations in \(\b_s\) \cite{HNP}. We provide a simple proof of this in \cref{sec:trans-ideal}.

While we have characterized the asymptotic symmetries as vector fields intrinsic to \(\scri\), they can also be obtained as limits to \(\scri\) of vector fields in the physical spacetime which preserve the asymptotic conditions in \cref{def:dna}. We give this alternative formulation in conformal Bondi-Sachs coordinates in \cref{sec:symm-BS}. 

Further, we have defined the asymptotic symmetries as infinitesimal diffeomorphisms that preserve the universal structure of \(\scri\). This leaves open the question of whether some of these symmetries are actually ``gauge'' in the sense of being degeneracies of the symplectic form of the theory (see \S\ IV of \cite{WZ} for a precise formulation of this condition). However, carrying out this symplectic analysis is complicated and one would also need to specify a Lagrangian for both the spacetime metric and the matter fields.  We hope that this can be addressed in a future analysis.

\begin{remark}[Comparison with the analysis of Kehagias-Riotto \cite{KR}]
\label{rem:KR-symm}
The above characterization of the asymptotic algebra is in contradiction with the claim of Kehagias and Riotto who find the BMS algebra for (linearly perturbed) FLRW spacetimes (\S~4 of \cite{KR}). Kehagias and Riotto start with the FLRW metric in the form \(\hat g_{ab} = a^2 \eta_{ab}\) with \(\eta_{ab}\) the Minkowski metric and consider infinitesimal transformations of the Minkowski metric  \(\Lie_\xi \eta_{ab}\) for some vector field \(\xi^a\) which is a BMS vector field in the Minkowski spacetime. However, infinitesimal diffeomorphism of FLRW and Minkowski metrics are related by
\be\label{eq:diff-FLRW-Mink}
    \Lie_\xi \hat g_{ab} = a^2 \Lie_\xi \eta_{ab} + 2(a \Lie_\xi a) \eta_{ab} = a^2 \Lie_\xi \eta_{ab} + 2 (a^{-1} \Lie_\xi a) \hat g_{ab} \; .
\ee
Thus, an infinitesimal diffeomorphism of FLRW cannot be written as \(a^2\) times an infinitesimal diffeomorphism of Minkowski metric unless \(\Lie_\xi a = 0\), i.e. \(\xi^a\) is a spatial translation or a rotation. Thus, apart from the exact Killing fields of FLRW spacetimes (see \cref{rem:FLRW-Killing} below), the asymptotic BMS symmetries derived in \cite{KR} \emph{do not} arise from any infinitesimal diffeomorphisms of the physical FLRW spacetimes. In other words, the infinitesimal diffeomorphisms in \cite{KR} are not related to any standard notion of asymptotic symmetries.
\end{remark}

\begin{remark}[Relation to the conformal Carroll algebra]
    The universal structure described above is a \emph{conformal Carroll structure} on \(\scri\). Carroll structures are obtained as the (degenerate) ultra-relativistic limit of Minkowski spacetime when the speed of light limits to zero. The symmetry groups of such structures were first studied by L\'evy-Leblond \cite{LL}. Including the freedom to perform conformal transformations then extends the Carroll group to the conformal Carroll group \cite{Duval:2014uva,Duval:2014lpa,Hartong:2015xda,Ciambelli:2019lap}. The asymptotic symmetry algebras \(\b_s\) obtained above are isomorphic to the Lie algebra of these conformal Carroll groups.
\end{remark}

\begin{remark}[Killing vector fields of decelerating FLRW spacetimes at \(\scri\)]
\label{rem:FLRW-Killing}
Since FLRW spacetimes are homogenous and isotropic, they have six Killing vector fields: three spatial translations $T^a_i$ and three rotations $R^a_i$ (with $i \in \{x,y,z\}$ representing the Cartesian directions in the spacelike surfaces of homogeneity). In the divergence-free conformal frame (see \cref{sec:conformalcompletionFLRW}) these take the following form at \(\scri\):
\begin{subequations}\begin{align}
T^a_i &=
	- 2^s (1-s) \begin{pmatrix}
		\sin \theta \cos \phi \\
		\sin \theta  \sin \phi \\
		\cos \theta
	\end{pmatrix}
	n^a \\
R^a_i &=
	\begin{pmatrix}
		- \sin \phi \\
		\cos \phi \\
		0
	\end{pmatrix}
	(\del_{\theta})^a +
	\begin{pmatrix}
		- \cot \theta \cos \phi \\
		- \cot \theta \sin \phi \\
		1
	\end{pmatrix}
	(\del_{\phi})^a \, ,
\end{align}\end{subequations}
where we use matrix notation to enumerate \(i \in \{x,y,z\}\). It is easy to verify that \(T^a_i\) and \(R^a_i\) are elements of the asymptotic symmetry algebra \(\b_s\), i.e., satisfy \cref{eq:conditionsxi} with $\alpha_{(T_i)} = \alpha_{(R_i)}=0$. The spatial translations are a subalgebra of the supertranslations \(\s_s\) (that is, they satisfy \cref{eq:Lieideal}) and the rotations are a subalgebra of the Lorentz algebra \(\mf{so}(1,3)\). As explained above (and proven in \cref{sec:trans-ideal}) in general there are no unique preferred translations in the algebra \(\b_s\); the spatial translations described above are picked uniquely by the exact FLRW spacetime. This is similar to how the BMS algebra does not contain a preferred Poincar\'e subalgebra but the exact Minkowski spacetime does pick a unique Poincar\'e subalgebra at null infinity.
\end{remark}

\begin{remark}[Topology of $\scri$ plays a key role]
The fact that the space of generators $S$ is topologically $\mathbb{S}^2$ is critical in the derivation of the BMS algebra. In the context of asymptotically flat spacetimes, \cite{Foster} investigated the possibility of relaxing this condition and found that depending on the global conformal structure of $S$ the asymptotic symmetry algebra can drastically change. In fact, generically no asymptotic symmetry algebra exists. However, to represent radiation from isolated sources in asymptotically flat spacetimes it is natural to require that $\scri$ is homeomorphic to $\mathbb{R} \times \mathbb{S}^2$. 
\end{remark}

\begin{remark}[Extensions of BMS]
It has been suggested that the BMS algebra in asymptotically flat spacetimes should be extended to include either the Virasoro algebra \cite{Barnich:2009se,Barnich:2010eb} or all diffeomorphisms of the \(2\)-sphere \cite{Campiglia:2014yka,CL}. However, the Virasoro vector fields are necessarily singular on the \(2\)-spheres and hence do not even preserve the smoothness structure of null infinity. Similarly, since the conformal class of the \(2\)-sphere metric is part of the universal structure, the extension by all diffeomorphisms of \(\bb S^2\) can always be reduced back to the Lorentz Lie algebra. It has also been shown that the extension to all diffeomorphisms of \(\bb S^2\) cannot be implemented in the covariant phase space in a local and covariant manner \cite{Flanagan:2019vbl}. Hence, for the remainder of this paper we shall not work with such enlarged symmetries.
\end{remark}

\section{Conformal and physical Bondi-Sachs coordinates}\label{sec:BS}

We have defined spacetimes with cosmological null asymptotes (\cref{def:dna}) in a covariant manner in terms of a conformal completion. But for certain practical applications, it is handy to have a suitable coordinate system at one's disposal. In this section, we construct such coordinates, in both the unphysical and physical spacetimes, similar to the Bondi-Sachs coordinates \cite{BBM,Sachs1} often used in the study of asymptotically flat spacetimes.\footnote{We emphasize that the Bondi-Sachs coordinates, though useful, are not ``divinely prescribed'' in any meaningful sense. One can construct other coordinates in both the unphysical and physical spacetimes that can be similarly useful. For instance, the conformal Gau{\ss}ian coordinates \cite{Hol-Th,HIW} near \(\scri\) are convenient in the study of asymptotic flatness in higher dimensions. These coordinates describing the unphysical spacetime are related to the affine-null coordinates of Winicour \cite{an-W} in the physical spacetime. Note that the coordinates used in \cref{sec:conformalcompletionFLRW} to construct the conformal completion of FLRW spacetimes are \emph{not} Bondi-Sachs coordinates.}\\

We first construct \emph{conformal Bondi-Sachs} coordinates for the unphysical metric in a neighbourhood of \(\scri\) using an asymptotic expansion. This asymptotic expansion also allows us to show, by direct computation, that the peeling theorem is not satisfied in this class of spacetimes (see \cref{rem:no-peeling}).

The conformal Bondi-Sachs coordinates can then be used to set up Bondi-Sachs-type coordinates, in which \(\scri\) is located at an ``infinite radial distance'' in the physical spacetime. Since --- unlike the asymptotically flat case --- the conformal factor \(\Omega\) is not smooth at \(\scri\) while \(\Omega^{1-s}\) is, we have two natural candidates for such a radial coordinate. We derive the asymptotic behaviour of the metric for both choices.

\subsection{Conformal Bondi-Sachs coordinates in the unphysical spacetime}
\label{sec:BS-unphys}

To construct the conformal Bondi-Sachs coordinates in a neighbourhood of \(\scri\), we first choose coordinates on \(\scri\) as follows. Let \(u\) be the parameter along the null generators so that \(n^a \grad_a u \hateq 1\), and let \(S_u \cong \bb S^2\) be the cross-sections of \(\scri\) with \(u = \text{constant}\). On some cross-section \(S_{u_0}\) with \(u \hateq u_0\) we pick coordinate functions\footnote{Note the precise choice of coordinates \(x^A\) on \(S_{u_0}\) is not relevant, one can choose polar coordinates, or stereographic coordinates or any other coordinates that one wishes. In general, we need more than one coordinate patch to cover all of \(S_{u_0} \cong \bb S^2\) but this subtlety will not be important.} \(x^A\) and parallel transport them to other cross-sections \(S_u\) along the null generators, \(n^a \grad_a x^A \hateq 0\). Then \((u, x^A)\) serve as coordinates on \(\scri\).

Next we need to pick a coordinate away from \(\scri\). Note that, unlike the asymptotically flat case, \(\grad_a \Omega \hateq 0\) for \(s \neq 0\) and so the conformal factor \(\Omega\) is not a ``good" coordinate away from \(\scri\). However, \(n_a = \tfrac{1}{1-s} \grad_a \Omega^{1-s} \not\hateq 0\) and thus we can use 
\begin{equation}
	\tilde \Omega \defn \tfrac{1}{1-s} \Omega^{1-s}
\end{equation}
as a coordinate function in a neighbourhood of \(\scri\) with \(\scri\) corresponding to \(\tilde\Omega = 0\).

Now we extend the coordinates \((u, x^A)\) away from \(\scri\). Consider the null hypersurfaces transverse to \(\scri\) that intersect \(\scri\) in the cross-sections \(S_u\). In a sufficiently small neighbourhood of \(\scri\), such null hypersurfaces do not intersect each other and thus generate a null foliation. We first extend the coordinate \(u\) by demanding that it be constant along  these null hypersurfaces. Then let \(l_a \defn - \nabla_a u\) be the future-directed null normal to these hypersurfaces --- i.e., \(l^al_a = 0\) --- normalised so that \(l^a n_a \hateq - 1\). Then, we extend the angular coordinates to a neighourhood of \(\scri\) by parallel transport \(l^a \nabla_a x^A = 0\). This concludes the setup of the \emph{conformal Bondi-Sachs coordinates} \((u,\tilde\Omega, x^A) \) in a neighbourhood of \(\scri\).\\

The general form of the unphysical metric in these coordinates is then
\be\label{eq:unphys-BS}
	ds^2 = - W e^{2\beta} du^2 + 2 e^{2\beta} du d\tilde\Omega + h_{AB} (dx^A - U^A du )(dx^B - U^B du ) \; ,
\ee
where \(W\), \(\beta\), \(h_{AB}\), and \(U^A\) are smooth functions of \((u, \tilde\Omega, x^A)\), and \(g_{\tilde\Omega \tilde\Omega} = g_{\tilde\Omega A} = 0\) follows from \(l^a l_a = l^a \nabla_a x^A = 0\).\\

The metric components are still rather generic. By making a particular choice for the conformal factor \(\Omega\), we restrict the freedom of the metric components \(h_{AB}\). As shown in \cref{sec:dna}, at \(\scri\) we can pick the conformal factor so that \(n^a\) is divergence-free and geodesic and we are in the Bondi conformal frame. In this frame, \(u\) is the affine parameter along the null generators. Further, the conformal factor can also be chosen such that on the cross-section \(S_{u_0}\), \(h_{AB} \hateq q_{AB}\) is the metric of a unit round sphere in the coordinates \(x^A\) (as discussed in \cref{sec:dna}, the coordinate invariant statement is that the Ricci scalar \(\ms R\) of the metric is \(2\)). Moreover, from \cref{eq:lie-n-q}, we have \(\partial_u q_{AB} \hateq 0\) and consequently the metric on \emph{all} cross-sections \(S_u\) is also the unit round metric. To pick the conformal factor away from \(\scri\), note that under an additional conformal transformation \(\Omega' = \omega \Omega \), with \(\omega\hateq 1\) we have
\be
	\det h' = \omega^4 \det h \; ,
\ee
where the determinant is computed in the choice of coordinates \(x^A\). Away from \(\scri\) we can use the above freedom in the conformal factor to impose
\be\label{eq:deth-cond}
\det h = \det q \; ,
\ee
that is, the spheres of constant \(u\) and \(\tilde\Omega\) have area \(4\pi\). This exhausts the freedom in the conformal factor. In the remainder of this section we will use the convention that the angular indices \(A,B,\ldots\) are raised and lowered with the unit round metric \(q_{AB}\) and denote the covariant derivative of \(q_{AB}\) by \(\eth_A\).

We assume that the metric components in \cref{eq:unphys-BS} have an asymptotic expansion in integer powers of \(\tilde\Omega\) near \(\scri\). In particular, we have
\be
	h_{AB} = q_{AB} + \tilde\Omega C_{AB} + \tilde\Omega^2 d_{AB} + \mc O(\tilde\Omega^3) \; .
\ee
Imposing \cref{eq:deth-cond} we get
\be\label{eq:trace-cond}
	q^{AB}C_{AB} = 0 \eqsp q^{AB} d_{AB} = \tfrac{1}{2} C^{AB} C_{AB} \; .
\ee
Similarly, for our choice of coordinates and in the Bondi conformal frame we have \(n^a \nabla_a u \hateq 1 \), \(n^a \nabla_a x^A \hateq 0\) and \(n_a n^a = \mc O(\tilde\Omega^2)\), which immediately imply the following falloff behavior of the expansion coefficients:
\be
	W = \tilde\Omega^2 W^{(2)} + \mathcal{O}(\tilde \Omega^3) \eqsp U^A = \tilde\Omega U^{(1)}{}^A + \mathcal{O}(\tilde \Omega^2)\eqsp \beta = \tilde\Omega \beta^{(1)} + \mathcal{O}(\tilde \Omega^2) \; .
\ee
Let us define \(\lim\limits_{\to \scri} 8\pi g^{ab}\hat T_{ab} \hateq \mc T  \) and decompose \(\tau_a\) as  
\be
\tau_a \hateq \tau n_a + \tau_A \nabla_a x^A \; ,
\ee 
where we used \cref{eq:n-tau} to set the $u$-component of $\tau_a$ to zero. Using \cref{eq:nisnull} we get
\be\label{eq:W2-soln}
	W^{(2)} = \tfrac{(1-s)^3}{2 (1-s^3)} \lb[ \mc T + 2 + 2 s \partial_u \tau + \tfrac{2s(2+s)}{1-s} \eth_A \tau^A  + \tfrac{1}{2} s^2 \tau_A \tau^A \rb] 
\ee
and similarly from \cref{eq:gradneqn} we obtain
\be\label{eq:U1-beta1-soln}
	U^A{}^{(1)} = s \tau^A \eqsp \beta^{(1)} = -\tfrac{1}{2} s \tau \; .
\ee
Note that the singular term in the stress-energy given by \(\tau_a\) appears in the leading-order non-trivial metric coefficients. For asymptotically flat spacetimes, $U^A{}^{(1)}$ and $\beta^{(1)}$ both vanish and $W^{(2)} = 1$.\\


To summarise, in the conformal Bondi-Sachs coordinates \((u,\tilde\Omega, x^A)\) we have the unphysical metric \cref{eq:unphys-BS} with the following asymptotic expansions
\begin{subequations}\label{eq:metric-BS-exp}\begin{align}
	W & = \tilde\Omega^2 W^{(2)} + \tilde\Omega^3 W^{(3)} + \mc O(\tilde\Omega^4) \\
	\beta & = -\tfrac{1}{2} s \tilde\Omega \tau + \tilde\Omega^2 \beta^{(2)} + \mc O(\tilde\Omega^3) \\
	U^A & = s \tilde\Omega \tau^A + \tilde\Omega^2 U^{(2)}{}^A + \tilde\Omega^3 U^{(3)}{}^A + \mc O(\tilde\Omega^4) \\
	h_{AB} & = q_{AB} + \tilde\Omega C_{AB} + \tilde\Omega^2 d_{AB} + \mc O(\tilde\Omega^3) \; ,
\end{align}\end{subequations}
where \(W^{(2)}\) is given in terms of the stress-energy tensor in \cref{eq:W2-soln}, $C_{AB}$ is traceless and the trace of $d_{AB}$ is specified in \cref{eq:trace-cond}. In the case of asymptotically flat spacetimes, the coefficients $W^{(3)}$ and $U^{(3)}{}^A$ are often written as
\be\label{eq:M-N-af}
    W^{(3)} \stackrel{\rm flat}{=} -2M \eqsp U^{(3)}{}^A \stackrel{\rm flat}{=} - \tfrac{2}{3} N^A + \tfrac{1}{16} \eth^A(C_{BC}C^{BC}) + \tfrac{1}{2} C^{AB} \eth^C C_{BC} \; ,
\ee
where $M$ and $N^A$ are referred to as the mass and angular momentum aspect, respectively \cite{BBM,Tam-Win}. Indeed, for asymptotically flat spacetimes \(M\) and \(N^A\) encode the mass and angular momentum of the spacetime at null infinity \cite{Flanagan:2015pxa}. This interpretation is supported by explicit examples such as the Kerr-Newman and Vaidya metric, the Landau-Lifschitz approach to defining balance laws \cite{Bonga:2018gzr} and put on a firm ground by an analysis of the covariant phase space of asymptotically flat spacetimes \cite{AshtekarBombelliReula, WZ,Flanagan:2015pxa}. Since a similar analysis has not been done for FLRW spacetimes, the coefficients $W^{(3)}$ and $ U^{(3)}{}^A$ should at this stage not be interpreted as the mass and angular momentum aspect in the usual sense; especially given that the asymptotic symmetry algebra does not have a preferred translation subalgebra --- as discussed in \cref{sec:symm,sec:trans-ideal} --- even the notion of mass appears to be ambiguous.

\begin{remark}[Failure of peeling]\label{rem:no-peeling}
    Since the unphysical metric \(g_{ab}\) is smooth at \(\scri\), so is its Weyl tensor. However, unlike the case of asymptotically flat spacetimes the Weyl tensor does not vanish at \(\scri\), \(C_{abcd} \not\hateq 0\), and consequently its decay is ``slower'' than that of asymptotically flat spacetimes. Specifically, at \(\scri\) we can choose the null tetrad 
\be
    n_a \hateq \nabla_a \tilde \Omega \eqsp l_a \hateq - \nabla_a u \eqsp m_a \hateq   \tfrac{1}{\sqrt{2}} (\nabla_a \theta + i \sin\theta \nabla_a\phi) \eqsp \bar m_a \hateq   \tfrac{1}{\sqrt{2}} (\nabla_a \theta - i \sin\theta \nabla_a\phi) \; ,
\ee
where \((\theta,\phi)\) are polar coordinates on the cross-sections of \(\scri\). Then, using \cref{eq:unphys-BS,eq:metric-BS-exp}, the Newman-Penrose components of the Weyl tensor at \(\scri\) are (following the conventions in \cite{SK})
\begin{subequations}\begin{align}
    \Psi_4 & \defn - C_{abcd} \bar{m}^a n^b \bar{m}^c n^d \hateq 0 \eqsp \tilde\Omega^{-1}\Psi_4 \hateq \lb( \tfrac{1}{2}\partial_u^2 C_{AB} + s \partial_u \eth_A \tau_B + \tfrac{1}{2} s^2 \tau_A \partial_u \tau_B \rb) \bar m^A \bar m^B \\
    \Psi_3 & \defn - C_{abcd} l^a n^b \bar{m}^c n^d\hateq - \tfrac{s}{4} \partial_u \tau_A \bar m^A \\
    \Psi_2& \defn  - C_{abcd} l^a m^b \bar{m}^c n^d \hateq  - \tfrac{1}{6} \lb[ W^{(2)}-1 - s \lb(\partial_u \tau + \tfrac{1}{2} \eth_A \tau^A + s \tau_A \tau^A + \tfrac{3}{2} i \epsilon^{AB} \eth_A \tau_B \rb) \rb]
\end{align}\end{subequations}
while \(\Psi_1\) and \(\Psi_0\) have more complicated expressions involving the second-order quantities \(d_{AB}\) and \(U^{(2)}{}^A\). From these expressions it is clear that the Weyl tensor does not respect the usual ``peeling'' order due to the presence of the matter terms encoded in $\tau_a$ and the deviation of $W^{(2)}$ from~1.
Note, however, that the Weyl tensor does vanish at \(\scri\) for the spacetimes constructed in \cref{sec:manyspacetimes} since their conformal completion is --- by construction --- the same as that of some asymptotically flat spacetime.
\end{remark}

\subsection{Bondi-Sachs-type coordinates in the physical spacetime}
\label{sec:BS-phys}

The conformal Bondi-Sachs coordinates for the unphysical spacetime constructed above can be used to obtain asymptotic coordinates for the physical metric. In the conformal Bondi-Sachs coordinates, the physical metric is
\be
	d\hat s^2  = \lb[ (1-s) \tilde\Omega \rb]^{-\tfrac{2}{1-s}} \bigg[ - W e^{2\beta} du^2 + 2 e^{2\beta} du d\tilde\Omega + h_{AB} (dx^A - U^A du)(dx^B - U^B du ) \bigg] \; .
\ee
Note that the surfaces of constant \(u\) are outgoing null surfaces in the physical spacetime. To put this metric in a more familiar form, we define a radial coordinate in the physical spacetime so that \(\scri\) is approached as the radial coordinate goes to infinity along the null surfaces of constant \(u\). There are two natural choices for such a radial coordinate, which we detail below. \\

Since, in the unphysical spacetime \(\tilde\Omega\) is a good coordinate at \(\scri\) where \(\tilde\Omega \hateq 0\), we can define the ``radial'' coordinate \(\tilde r\) by
\be
	\tilde r \defn \tilde\Omega^{-1} 
\ee
so that \(\scri\) is approached as \(\tilde r \to \infty\). In these coordinates \((u, \tilde r, x^A)\), the physical metric is
\be\label{eq:phys-metric-tilde}
	d\hat s^2 = \lb(\frac{\tilde r^s}{1-s}\rb)^{\tfrac{2}{1-s}} \bigg[ - \frac{\tilde V}{\tilde r}  e^{2\beta} du^2 - 2 e^{2\beta} du d\tilde r + \tilde r^2 h_{AB} (dx^A - U^A du)(dx^B - U^B du ) \bigg] \; ,
\ee
where \(\tilde V \defn \tilde r^3 W\). The overall ``scale factor" \(\tilde r^{\tfrac{2s}{1-s}}\) does not fall off as \(\tilde r \to \infty\), but the remaining metric components have the following falloffs in integer powers of \(1/\tilde r\)
\begin{subequations}\label{eq:phys-falloff-tilde}\begin{align}
	\frac{\tilde V}{\tilde r} & = W^{(2)} + \frac{W^{(3)}}{\tilde r} + \mc O(1/\tilde r^2) \\
	e^{2\beta} & = 1 - \frac{s \tau}{\tilde r} + \frac{1}{\tilde r^2} \lb( 2 \beta^{(2)} + \tfrac{1}{2} s^2 \tau^2 \rb) + \mc O(1/\tilde r^3) \\
	U^A & = \frac{1}{\tilde r} s \tau^A + \frac{1}{\tilde r^2} U^{(2)}{^A} + \frac{U^{(3)}{}^A}{\tilde r^3} + \mc O(1/\tilde r^4) \\
	h_{AB} & = q_{AB} + \frac{1}{\tilde r} C_{AB} + \frac{1}{\tilde r^2} d_{AB} + \mc O(1/\tilde r^3) \; .
\end{align}\end{subequations}
In these coordinates the physical area of the spheres of constant \(u\) and \(\tilde r\) is \(4\pi \lb( \frac{\tilde r}{1-s} \rb)^{\tfrac{2}{1-s}}\).\\

We can instead follow the original construction by Bondi and van der Burg (Part B of \cite{BBM}) and define a radial coordinate \(r\) so that the physical area of the spheres of constant \(u\) and \(r\) is \(4\pi r^2\). In terms of the conformal factor this is given by (see also \cite{Tam-Win})
\be
	r \defn \Omega^{-1} = \lb( \frac{\tilde r}{1-s} \rb)^{1/(1-s)}
\ee
and the physical metric is
\be\label{eq:phys-metric-BS}
	d\hat s^2 = - \frac{V}{r} e^{2\beta} du^2 - 2 r^s e^{2\beta} du dr + r^2 h_{AB} (dx^A - U^A du)(dx^B - U^B du ) 
\ee
with \(V \defn r^3 W\). However, note that the metric components now do not have an expansion in terms of integer powers of \(1/r\) when \(s \neq 0\), instead we have
\begin{subequations}\label{eq:phys-falloff-BS}\begin{align}
	\frac{V}{r} & = \frac{r^{2s}}{(1-s)^2}\bigg[ W^{(2)} - \frac{1}{(1-s)}\frac{W^{(3)}}{r^{1-s}} + \mc O(1/r^{2(1-s)}) \bigg] \\
	r^s e^{2\beta} & = r^s \lb[ 1 - \frac{1}{(1-s)}\frac{s \tau}{r^{1-s}} + \frac{1}{(1-s)^2}\frac{1}{r^{2(1-s)}} \lb( 2 \beta^{(2)} + \tfrac{1}{2} s^2 \tau^2 \rb) + \mc O(1/r^{3(1-s)}) \rb]\\
	U^A & =  \frac{1}{(1-s)}\frac{1}{r^{1-s}} s \tau^A + \frac{1}{(1-s)^2}\frac{1}{r^{2(1-s)}} U^{(2)}{^A} + \frac{1}{(1-s)^3}\frac{U^{(3)}{}^A}{r^{3(1-s)}} + \mc O(1/r^{4(1-s)}) \\
	h_{AB} & = q_{AB} + \frac{1}{(1-s)}\frac{1}{r^{1-s}} C_{AB} + \frac{1}{(1-s)^2}\frac{1}{r^{2(1-s)}} d_{AB} + \mc O(1/r^{3(1-s)}) \; .
\end{align}\end{subequations}\\


In the asymptotically flat case, both choices for the radial coordinate are the same  \(\tilde r = r\) and we reproduce the usual falloff conditions in Bondi-Sachs coordinates (see for example \cite{Flanagan:2015pxa}). Just as in the asymptotically flat case, one can use the falloffs in either \cref{eq:phys-metric-tilde,eq:phys-falloff-tilde} or \cref{eq:phys-metric-BS,eq:phys-falloff-BS} as the definition of the asymptotics of the spacetimes in \cref{def:dna} in these choices of coordinates.

\begin{remark}
\label{rem:coordinate-fall-off}
Note that neither \(\tilde r\) nor \(r\) is a well-behaved coordinate at \(\scri\). In particular, the coordinate basis covectors
\be
	(d\tilde r)_a = - \tilde\Omega^{-2} n_a \eqsp (dr)_a = - \Omega^{s-2} n_a
\ee
diverge at \(\scri\) and, similarly, the coordinate basis vectors \((\del_{\tilde r})^a\) and \((\del_r)^a\) vanish at \(\scri\). Thus, one must be careful while dealing with \(\tilde r\)- or \(r\)-components of tensors. For instance, consider the leading divergent piece of the stress-energy tensor \(\hat T_{ab}\) in \cref{eq:T-exp}, which in these coordinates takes the form
\be
	2 s \Omega^{2(s-1)} n_a n_b = \frac{2 s}{(1-s)^2 \tilde r^2} (d\tilde r)_a (d\tilde r)_b = \frac{2s}{r^2} (dr)_a (dr)_b \; .
\ee 
Thus, the components \(\hat T_{\tilde r \tilde r}\) and \(\hat T_{rr}\) fall off as \(1/\tilde r^2\) and \(1/r^2\), respectively, near \(\scri\) (these are the falloff conditions imposed in Eq.~A.4 of \cite{KR}). However, one \emph{should not} conclude that this stress-energy tensor is smooth at, or decays towards, \(\scri\). A \(\hat T_{ab}\) which is smooth at \(\scri\) falls off as \(\hat T_{\tilde r \tilde r} \sim 1/\tilde r^4\) and \(\hat T_{rr} \sim 1/r^4\), and for an asymptotically flat stress-energy (\(\Omega^{-2}\hat T_{ab}\) being finite at \(\scri\)) we have \(\hat T_{rr} \sim 1/r^6\). 
\end{remark}

\subsection{FLRW in Bondi-Sachs coordinates}
\label{sec:FLRW-BS}

FLRW spacetimes being our canonical example for spacetimes satisfying \cref{def:dna}, in this section we compute the expansion coefficients of the metric components of decelerating FLRW spacetimes in the physical \((u, \tilde r, x^A)\) Bondi-Sachs coordinates constructed above. The computation below is in essence an explicit version of the construction in \cref{sec:manyspacetimes}.

Consider the physical FLRW metric
\be\label{eq:FLRW-Mink-BS}\begin{aligned}
    d\hat s^2 & = a^2 ( -d\eta^2 + d \af r^2 + \af r^2 q_{AB}d\af x^A d\af x^B ) \\
    & = a^2 (- d\af u^2 - 2 d\af u d \af r + \af r^2 q_{AB}d\af x^A d\af x^B ) \; ,
\end{aligned}\ee
where in the second line we have switched to a Bondi-Sachs coordinate system for the flat spacetime, with \(\af u \defn \eta - \af r\) and the FLRW scale factor is given by
\be\label{eq:scale-Mink-BS}
    a^2 = (\af r + \af u)^{2s/(1-s)} = \af r^{2s/(1-s)} A^{2s/(1-s)} \eqsp A = 1 + \frac{\af u}{\af r} \; .
\ee
Apart from the overall \(\af r^{2s/(1-s)}\), we can expand \(A^{2s/(1-s)}\) in inverse powers of \(\af r\) to get the asymptotic form of the metric for large \(\af r\).  However, this asymptotic form of the metric does not coincide with the one given in \cref{eq:phys-metric-tilde,eq:phys-falloff-tilde}, since the coordinates are adapted to the flat and not the FLRW spacetime. This is illustrated by the fact that the area of spheres with constant \(\af u\) and \(\af r\) depend on the values of \(\af u\) and \(\af r\).

To put the FLRW metric in the form given in \cref{eq:phys-metric-tilde,eq:phys-falloff-tilde} we choose new coordinates \(( u, \tilde r, x^A)\) defined by
\be\label{eq:phys-FLRW-coords-tilde}
    u = (1-s) \af u \eqsp \tilde r = (1-s) \af r A^s \eqsp x^A = \af x^A \; .
\ee
In these coordinates the scale factor is given by
\be\label{eq:scale-FLRW-tilde}
    a^2 = \lb( \frac{\tilde r^s}{1-s} \rb)^{2/(1-s)} (1-s)^2 A^{2s}
\ee
and, using \cref{eq:scale-Mink-BS,eq:phys-FLRW-coords-tilde}, the function \(A\) satisfies the equation
\be
    A = 1+ A^s \frac{u}{\tilde r} \; .
\ee
For large \(\tilde r\), this has the solution
\be\label{eq:A-soln}
    A = 1 + \frac{u}{\tilde r} + s \frac{u^2}{\tilde r^2} + \mc O(1/\tilde r^3) \; .
\ee

In the new coordinates \cref{eq:phys-FLRW-coords-tilde}, the FLRW metric in \cref{eq:FLRW-Mink-BS} takes the form in \cref{eq:phys-metric-tilde} with the metric components falling off as
\be\begin{aligned}
    \frac{\tilde V}{\tilde r} &= (1-2s) + 2 s(1-s) \frac{u}{\tilde r} - \frac{3}{2} s (1-s) (1-2s) \frac{u^2}{\tilde r^2} + \mc O(1/\tilde r^3) \\
    e^{2\beta} &= 1 + 2s \frac{u}{\tilde r} - \frac{3}{2} s (1-3s) \frac{u^2}{\tilde r^2} + \mc O(1/\tilde r^3) \\
    h_{AB} &= q_{AB} \eqsp U^A = 0 \; ,
\end{aligned}\ee
where we used \cref{eq:scale-FLRW-tilde,eq:A-soln}.
Note that since an FLRW spacetime is isotropic (spherically symmetric) and the coordinate system \((u, \tilde r, x^A)\) is adapted to this spherical symmetry, the metric coefficients \(U^A\) vanish to all orders. Further, for the same reason all coefficients in the \(1/\tilde r\) expansion of \(\tilde V/\tilde r\) and \(e^{2\beta}\) are functions of  \(u\) only. However, the homogeneity (spatial translation symmetry) of the FLRW spacetime is not manifest in these coordinates.

\subsection{Asymptotic symmetries in Bondi-Sachs coordinates}\label{sec:symm-BS} 

In \cref{sec:symm} we described the asymptotic symmetries of spacetimes with a cosmological null asymptote in terms of vector fields intrinsic to \(\scri\) that preserve the universal structure. Here, we will briefly describe these asymptotic symmetries in terms of vector fields in the conformal Bondi-Sachs coordinates constructed above.\footnote{A similar description can also be given in terms of the physical coordinates following the analysis of \cite{Sachs2}.}\\

Let \(\xi^a\) be any vector field on \(\hat M\) and let \(\hat\xi_a = \hat g_{ab} \xi^b\). 
The physical metric perturbation generated by a diffeomorphism along \(\xi^a\) is \(\hat\gamma_{ab} = \Lie_\xi \hat g_{ab} = 2 \hat\nabla_{(a} \hat\xi_{b)} \). 
Let \(\xi^a\) extend smoothly to the conformally completed spacetime \(M\) --- so that it preserves the smooth differential structure at \(\scri\), and denote the covector by \(\xi_a = g_{ab} \xi^b = \Omega^2 \hat\xi_a\). The corresponding perturbation to the conformally completed metric \(g_{ab}\) is given by 
\be\label{eq:unphys-diff}\begin{split}
	\gamma_{ab} & = \Omega^2 \hat\gamma_{ab} = 2 \nabla_{(a} \xi_{b)} - 2 \Omega^{-1} \nabla_c \Omega~ \xi^c g_{ab} \\
	& = 2 \nabla_{(a} \xi_{b)} - \tfrac{2}{1-s} \tilde\Omega^{-1} \,  n_c \xi^c g_{ab} \; .
\end{split}\ee
Preserving the smoothness of \(g_{ab}\) at \(\scri\) requires that \(\gamma_{ab}\) also be smooth at \(\scri\). Consequently, \cref{eq:unphys-diff} implies that $\xi^a n_a \hateq 0$; $\xi^a$ is tangential to $\scri$.

To extract more information from the above equation, let us expand the components of $\xi^a$ also in powers of $\tilde{\Omega}$ (note \(\xi^{\tilde\Omega}_{(0)} = 0\) since \(\xi^a n_a \hateq 0\))
\be\begin{split}
	\xi^u & = F + \tilde\Omega \xi^u_{(1)} + \tilde\Omega^2 \xi^u_{(2)} + \mc O(\tilde\Omega^3) \\
	\xi^{\tilde\Omega} & = \tilde\Omega \xi^{\tilde\Omega}_{(1)} + \tilde\Omega^2 \xi^{\tilde\Omega}_{(2)} + \mc O(\tilde\Omega^3) \\
	\xi^A & = X^A + \tilde\Omega \xi^A_{(1)} + \tilde\Omega^2 \xi^A_{(2)} + \mc O(\tilde\Omega^3) 
\end{split}
\label{eq:falloff-vectorfields}
\ee
and then impose the falloff conditions \cref{eq:unphys-BS,eq:metric-BS-exp} on \(\gamma_{ab}\) order by order in $\tilde{\Omega}$.
At leading order, i.e., from the vanishing of \(\gamma_{ab}\) at order \(\tilde\Omega^0\), we obtain the following set of equations
\begin{subequations}\label{eq:diffeo-0}\begin{align}
	\partial_u F & = \tfrac{1+s}{1-s} \xi^{\tilde\Omega}_{(1)} \\
	\xi^u_{(1)} & = \partial_u X^A  = 0 \\
	\xi^A_{(1)} & = - \eth^A F \\
	2 \eth_{(A} X_{B)} & = \tfrac{2}{1-s} q_{AB} \xi^{\tilde\Omega}_{(1)} \; .
\end{align}\end{subequations}
Furthermore, by contracting \cref{eq:unphys-diff} with \(g^{ab}\) we find that $\tilde\Omega^{-1}  n_c \xi^c \hateq \tfrac{1-s}{4} \grad_a \xi^a$, which imposes the following additional relation:
\be
	- \tfrac{3+s}{1-s} \xi^{\tilde\Omega}_{(1)} + \eth_A X^A + \partial_u F = 0 \; .
\ee
Solving these conditions, we find that the order \(\tilde\Omega^0\) parts of $\xi^a$ satisfy
\be\label{eq:symm-BS}\begin{split}
	& 2 \eth_{(A} X_{B)} = q_{AB} \eth_C X^C \eqsp F = f + \tfrac{1+s}{2} u \eth_A X^A \quad \text{with } \quad \partial_u f = \partial_u X^A = 0 \; .
\end{split}\ee
This illustrates that the vector fields which preserve the falloff conditions \cref{eq:unphys-BS,eq:metric-BS-exp} in the conformal Bondi-Sachs coordinates induce the vector field \(f\partial_u + X^A \partial_A\) on $\scri$, where \(f(x^A)\del_u \) is a supertranslation and \(X^A\) a conformal Killing field on \(\bb S^2\), with \(\alpha_{(f)} = 0\) and \(\alpha_{(X)} = \tfrac{1}{2} \eth_A X^A\) (in \cref{eq:conditionsxi}). Explicitly evaluating the Lie bracket between these vector fields, one can confirm that \cref{eq:Lieideal} is satisfied. Thus, we recover the asymptotic symmetry algebra \(\b_s\) discussed in \cref{sec:symm}. Note that the dependence on $s$ in \cref{eq:symm-BS} also implies that this algebra is not isomorphic to the BMS algebra.

From \cref{eq:diffeo-0}, one can also solve for the order \(\tilde\Omega^1\) part of these asymptotic symmetry vector fields:
\be\label{eq:diffeo-1}
	\xi^u_{(1)} = 0 \eqsp \xi^{\tilde\Omega}_{(1)} = \tfrac{1-s}{2} \eth_A X^A \eqsp \xi^A_{(1)} = - \eth^A F \; .
\ee
Evaluating \(\gamma_{ab}\) at order \(\tilde\Omega^1\) we get the transformations of the coefficients of the metric $\beta^{(1)}, U_A^{(1)}$ and $C_{AB}$. To determine how these coefficients transform, we need to determine $\xi^{\tilde\Omega}_{(2)}$, which can be obtained by preserving the condition \(q^{AB}C_{AB} = 0\) under diffeomorphisms:%
\footnote{For completeness, we also include the other components of $\xi^a$ at order \(\tilde\Omega^2\), although these are not used in this paper: \(\xi^u_{(2)} = 0 \eqsp \xi^A_{(2)} = \tfrac{1}{2} C^A{}_B \eth^B F - \beta^{(1)} \eth^A F \).}
\be\begin{split}
	q^{AB} \delta_\xi C_{AB} & = q^{AB} \Lie_X C_{AB} + 2 \eth^A \xi_A^{(1)} - 2 U_A^{(1)} \eth^A F - \tfrac{4}{1-s} \xi^{\tilde\Omega}_{(2)} = 0 \\
	\implies \xi^{\tilde\Omega}_{(2)}  & = - \tfrac{1-s}{2} \lb[ \eth^2 F + U_A^{(1)} \eth^A F \rb] \; ,
\end{split}\ee
where we have used \cref{eq:diffeo-1} and that \(X^A\) is a conformal Killing field for \(q_{AB}\). Then we have
\begin{subequations}\begin{align}
	\delta_\xi \beta^{(1)} & = F \partial_u \beta^{(1)} + \Lie_X \beta^{(1)} + \tfrac{1-s}{2} \beta^{(1)} \eth_A X^A + \tfrac{1+s}{2} U_A^{(1)} \eth^A F + \tfrac{s}{2} \eth^2 F
	\\
	\delta_\xi U_A^{(1)} & = F \partial_u U_A^{(1)} + \Lie_X U_A^{(1)} + s \eth_A \eth_B X^B \\
	\begin{split}
	\delta_\xi C_{AB} & = F\partial_u  C_{AB} - 2 \lb( \eth_A \eth_B F - \tfrac{1}{2} q_{AB} \eth^2 F \rb) + \Lie_X C_{AB} - \tfrac{1+s}{2} C_{AB} \eth_C X^C \\
	&\quad - 2 \lb( U_{(A}^{(1)} \eth_{B)} F - \tfrac{1}{2} q_{AB} U_C^{(1)} \eth^C F \rb) \; .
	\end{split}
\end{align}\end{subequations}
These expressions reduce to their asymptotically flat counterparts when $s=0$. From \cref{eq:U1-beta1-soln}, we can rewrite the first two equations in terms of \(\tau_a\) (the diverging piece of the stress-energy) as 
\begin{subequations}\begin{align}
	\delta_\xi \tau & = F \partial_u \tau - (1+s) \tau_A \eth^A F - \eth^2 F + \Lie_X \tau + \tfrac{1-s}{2} \tau \eth_A X^A  \\
	\delta_\xi \tau_A & = F \partial_u \tau_A + \Lie_X \tau_A + \eth_A \eth_B X^B \; .
\end{align}\end{subequations}
The higher order coefficients of \(\xi^a\) and transformations of the metric coefficients can be obtained in an analogous way.

\section{Discussion}
\label{sec:dis}

We considered the structure at null infinity \(\scri\) of decelerating, spatially flat FLRW spacetimes. While the conformal completion of these spacetimes looks similar to that of Minkowski, we pointed out two crucial differences:
\begin{enumerate*}
    \item The conformal factor \(\Omega\) is not smooth at \(\scri\) and its gradient does not define a ``good'' normal to \(\scri\) since it vanishes there;
    \item The stress-energy tensor does not decay in the limit to null infinity, and in fact diverges.
\end{enumerate*}
These differences can be characterized by a parameter \(s\) --- related to the deceleration parameter and the equation of state as in \cref{eq:s-w-q} --- where \(0 \leq s < 1\) and \(s=0\) being Minkowski spacetime.

With this structure in mind, we defined a class of spacetimes with a cosmological null asymptote (\cref{def:dna}) whose behaviour at null infinity is similar to FLRW spacetimes instead of (asymptotically) flat spacetimes. We showed that the universal structure within this class of spacetimes is determined by an equivalence class of conformally related pairs $(q_{ab},n^a)$, where the difference with asymptotically flat spacetimes shows up in the conformal transformation of the normal $n^a \mapsto \omega^{-1-s} n^a$. As a result, the asymptotic symmetry algebra, i.e., the Lie algebra of vector fields preserving the universal structure, is very similar to the BMS algebra, but not isomorphic to it (contrary to the claim in \cite{KR}). The asymptotic symmetry algebra still has the structure of a semi-direct product of supertranslations with the Lorentz algebra, but the Lie bracket between a supertranslation and a Lorentz generator is now $s$-dependent (see \cref{eq:Lieideal}). As a consequence, the asymptotic symmetry algebra does not have any preferred translation subalgebra whenever $s\neq 0$.

Given the historical importance and widespread use of Bondi-Sachs coordinates, we also constructed Bondi-Sachs-like coordinates for spacetimes with a cosmological null asymptote in \cref{sec:BS}. We have done this for both the unphysical and physical metric, where for the latter we have used two different natural choices for the radial coordinate. Using the coordinates for the unphysical metric, the conformal Bondi-Sachs coordinates, we also showed that the peeling theorem for the components of the Weyl tensor does not apply for this class of spacetimes.\\

In this paper, we have focused on the universal structure of spacetimes with a cosmological null asymptote with the goal of finding the asymptotic symmetry algebra. In asymptotically flat spacetimes, the gravitational radiation is encoded in the next-order structure \cite{Ashtekar:1981hw}. Studying the next-order structure in this class of spacetimes with a cosmological null asymptote is a natural extension. Given that the canonical examples of this class of spacetimes are FLRW spacetimes, which are not stationary, the distinction between radiation and expansion is likely more subtle than for asymptotically flat spacetimes. 
There is some recent progress in understanding gravitational radiation emitted by compact sources in cosmological spacetimes, but within the context of linear perturbation theory \cite{Chu:2011ip,Chu:2015yua,Chu:2016ngc,Chu:2016qxp,Chu:2020sdn}.
Whether a nonlinear characterization of gravitational radiation can be achieved in the class of spacetimes derived here and its relation to possible observations by future gravitational wave detectors remains an open question.

Another closely related issue is whether the structure at null infinity of spacetimes with a cosmological null asymptote is stable under linear perturbations. This, and its mathematical big brother --- showing that there is a large class of initial data that have a cosmological null asymptote --- would be interesting for future research.

Additionally, a connected problem is the study of the memory effect in this class of spacetimes and its relation to the asymptotic symmetry algebra. In perturbed FLRW spacetimes, the memory effect has been studied using a local definition of memory \cite{TW}.\footnote{Others have also studied the memory effect in cosmological spacetimes  \cite{BGY,BGYu}. However, the class of spacetimes of interest to those authors involved a positive cosmological constant and are thus expanding in an accelerated fashion. Those spacetimes are not included in the class of spacetimes satisfying \cref{def:dna}.} In that work,  memory was associated with the derivative of a delta function in the linearized Riemann tensor of a retarded wave solution and did not involve any limits to $\scri$. As a result, no explicit connection to the asymptotic symmetry algebra was made.\\

As we noted in \cref{sec:symm}, we defined the asymptotic symmetry algebra as those vector fields that preserve the universal structure at \(\scri\). We have not addressed whether some of these symmetries are degeneracies of the symplectic form of the theory. Our analysis only used the asymptotic behaviour of the metric and the stress-energy tensor at \(\scri\) without specifying any particular Lagrangian for the matter fields of the theory. If one chooses a Lagrangian for the matter fields one should be able to carry out the symplectic analysis following the procedure described in \cite{WZ}. We leave this for future work. 

It would also be interesting to see if one can define the charges and their fluxes through null infinity similar to those in the asymptotically flat case \cite{BBM, Sachs1, Sachs2, Penrose, Geroch-asymp, GW, Ashtekar:1981bq, WZ}. However, since the symmetry algebra does not have any preferred translation subalgebra it seems that the notion of mass and linear-momentum in this class of spacetime is ambiguous. Finally, an active area of research today is the study of balance laws from past null infinity to future null infinity in the context of asymptotically flat spacetimes \cite{CE,KP-EM-match,Tro,KP-GR-match}. The canonical examples of spacetimes with a cosmological null asymptote are FLRW spacetimes, which only have either a past or future null infinity. Hence, the problem of balance laws might not have a direct analogue in this context. What does remain interesting is to connect the structure on null infinity to that at future timelike infinity $i^+$ \cite{Shiromizu:1999iq}. For instance, any definition of mass at null infinity should match that at future timelike infinity.

In \cref{sec:open}, we briefly summarized the conformal completions of spatially open FLRW universes which are not considered in the main paper. Since the behaviour of such spacetimes at null infinity is very distinct from that of spatially flat spacetimes, it would be of interest to investigate this case further.

\section*{Acknowledgements}
B.B. would like to thank Abhay Ashtekar, David Garfinkle and Robert M. Wald for stimulating discussions during the ``Mass in General Relativity'' Workshop at the Simons Center for Geometry and Physics and the APS April meeting in Columbus, Ohio. 
K.P. is supported in part by the NSF grant PHY-1801805.
This work was supported in part by the NSF grants PHY-1404105 and PHY-1707800 to Cornell University.
This research was supported in part by Perimeter Institute for Theoretical Physics. Research at Perimeter Institute is supported by the Government of Canada through the Department of Innovation, Science and Economic Development Canada and by the Province of Ontario through the Ministry of Research, Innovation and Science.
Some calculations used the computer algebra system \textsc{Mathematica} \cite{Mathematica}.

\appendix

\section{The algebra \(\b_s\) does not contain any preferred translation subalgebra}
\label{sec:trans-ideal}

In this Appendix, we will show explicitly that the asymptotic symmetry algebra \(\b_s\) does not contain any preferred translation subalgebra. In order to do so, we consider the Lie bracket of a supertranslation \(f n^a \in \s_s\) and a Lorentz vector field \(X^a\)  (see \cref{eq:Lieideal} with \(\alpha_{(X)} = \tfrac{1}{2} \eth_a X^a\))
\be\label{eq:beta-defn}
     [X, f n]^a  = \left[X^b \eth_b f - \tfrac{1}{2}(1+s) \eth_b X^b f \right] n^a \equiv F n^a\; .
\ee
The Lorentz vector field \(X^a\) is a vector field on \(\bb S^2\) and $F$ is defined to be the part between the square brackets and should not be confused with the leading order component of $\xi^u$ in \cref{eq:falloff-vectorfields}.   
If translations are a Lie ideal in \(\b_s\) then \(F\) would also be a translation whenever \(f\) is a translation. 
If \(f\) is a translation, \(f\) is a \(\ell = 0,1\) spherical harmonic on \(\bb S^2\). So if translations are a Lie ideal, \(F\) should also be a $L=0,1$ spherical harmonic on $\bb S^2$ whenever $\ell=0,1$. We will show explicitly that this is not the case.

First, decompose \(X^a\) into an ``electric'' and ``magnetic'' part (or equivalently, into a ``parity-even'' and ``parity-odd'' part)
\be\label{eq:X-decomp}
    X^a = \eth^a \beta + \epsilon^{ab} \eth_b \rho 
\ee
for some functions \(\beta\) and \(\rho\) which are \(\ell'\)-spherical harmonics. Since \(X^a\) is an element of the Lorentz algebra \(\mf{so}(1,3)\), both \(\beta\) and \(\rho\) are spherical harmonics with \(\ell' = 1\) (see \cite{Flanagan:2015pxa}). The function \(\beta\) corresponds to Lorentz boosts while \(\rho\) corresponds to Lorentz rotations.
Using the decomposition \cref{eq:X-decomp} in \cref{eq:beta-defn} we have
\be\
    F = \eth^a \beta \eth_a f - \tfrac{1}{2} (1+s) \eth^2 \beta  f - \epsilon^{ab} \eth_a \rho~ \eth_b f \; .
\ee
Now we wish to find the spherical harmonic mode \(L\) of \(F\) when \(f\) is a translation i.e. a \(\ell = 0,1\)-harmonic mode while the harmonic mode of \(\beta\) and \(\rho\) is \(\ell' = 1\). It is useful to consider the following different cases.
\paragraph*{Case 1: \(f\) is time translation \((\ell = 0)\).}
Whenever $f$ is a time translation, $F$ is either zero or a $L=1$ spherical harmonic and therefore this case does not does challenge the existence of a translation subalgebra in $\b_s$:
\be
    F = - \tfrac{1}{2} (1+s) \eth^2 \beta \; f = \tfrac{1}{2} (1+s) \ell'(\ell'+1) \beta \; f
\ee
so that
\be
    \eth^2 F = - \ell' (\ell' + 1) F = - L(L+1)F \; .
\ee
Hence, \(F = 0\) if \(\beta=0\) else \(F\) is a \(L = \ell' = 1\) mode. This implies that time translations are invariant under Lorentz rotations given by \(\rho\) but changes by a spatial translation under Lorentz boosts given by \(\beta\).
 
\paragraph*{Case 2: \(f\) is spatial translation \((\ell =  1)\) and \(X^a\) is a Lorentz rotation (\(\beta=0\) and \(\rho \neq 0\)).} 
This case also does not spoil the existence of a translation subalgebra. In particular, we have
\be
    F = - \epsilon^{ab} \eth_a \rho \eth_b f
\ee
and consequently
\be\begin{aligned}
    \eth^2 F & = \lb[ - \ell'(\ell'+1) - \ell(\ell+1) + 2 \rb] F - 2 \epsilon^{ab} \eth_c \eth_a \rho ~ \eth^c \eth_b f \\
    & = - \ell' (\ell' + 1) F = - L(L+1)F \; ,
\end{aligned}\ee
where in the first line we used that the Riemann tensor on $\bb S^2$ is $\ms R_{abcd} = q_{ac} q_{bd} - q_{ad}q_{bc}$ and in the last line we used that \(\ell=1\) and \(\eth_a \eth_b f = - q_{ab} f \) for $\ell=1$ spherical harmonics. Thus, \(F\) is a \(L = \ell' = 1\) mode. This means that a spatial translation changes by another spatial translation under Lorentz rotations.

\paragraph*{Case 3: \(f\) is spatial translation \((\ell =  1)\), and \(X^a\) is a Lorentz boost (\(\beta\neq 0\) and \(\rho=0\)).} 
In this case, we have
\be
    F = \eth^a \beta \eth_a f - \tfrac{1}{2} (1+s) \eth^2 \beta \; f \; .
\ee
To find the \(L\)-mode of \(F\), we multiply the above equation with the (complex conjugate) spherical harmonic \(\bar Y_{L,M}\) and integrate over \(\bb S^2\) to get
\be
    \int F \bar Y_{L,M} = \int \eth^a \beta ~ \eth_a f\;  \bar Y_{L,M} + \tfrac{1}{2}(1+s) \ell' (\ell'+1) \int \beta \; f \bar Y_{L,M} \; ,
\ee
where we have left the area element of the unit-metric on \(\bb S^2\) implicit for notational convenience.
The first term on the right-hand-side can be rewritten using repeated integration-by-parts as
\be\begin{aligned}
    \int \eth^a \beta \eth_a f \; \bar Y_{L,M}
    & = - \int \beta \; \eth^2 f \; \bar Y_{L,M} - \int \beta \; \eth_a f \eth^a \bar Y_{L,M} \\
    & =- \int \beta \; \eth^2 f \; \bar Y_{L,M}  + \int  \eth_a \beta f \eth^a \bar Y_{L,M} + \int \beta f \eth^2 \bar Y_{L,M} \\
    & = - \int \beta \; \eth^2 f \; \bar Y_{L,M} - \int  \eth^2 \beta f \bar Y_{L,M} + \int \beta f \eth^2 \bar Y_{L,M} - \int \eth_a \beta \eth^a f \; \bar Y_{L,M} \\
   	\implies \int \eth^a \beta \eth_a f \bar Y_{L,M} 
     & = - \tfrac{1}{2} \int \beta \; \eth^2 f \; \bar Y_{L,M}  - \tfrac{1}{2} \int  \eth^2 \beta \; f \; \bar Y_{L,M} + \tfrac{1}{2} \int \beta f \; \eth^2 \bar Y_{L,M} \\
    & = \tfrac{1}{2} \lb[ \ell(\ell+1) + \ell'(\ell'+1) - L(L+1) \rb] \int \beta f \; \bar Y_{L,M} \; .
\end{aligned}\ee
Thus, we obtain
\be
    \int F\;  \bar Y_{L,M} = \tfrac{1}{2} \lb[ \ell(\ell+1) + (2+s) \ell'(\ell'+1) - L(L+1) \rb] \int \beta f \; \bar Y_{L,M} \; .
\ee
Expanding the functions \(\beta\) and \(f\) in terms of the corresponding spherical harmonics \(Y_{\ell'=1, m'}\) and \(Y_{\ell=1,m}\), respectively, we can write the final integral in terms of the \(3\-j\)-symbols (see \S~34 of \cite{DLMF}) 
\be
    \int F \; \bar Y_{L,M} \propto  \lb[ 3+s - \tfrac{1}{2} L(L+1) \rb] 
    \begin{pmatrix}
    1 \; & 1 \; & L \\
    0 & 0 & 0 
    \end{pmatrix}
    \begin{pmatrix}
    1 & 1 & L \\
    m & m' & -M 
    \end{pmatrix} \; ,
\ee
where we have dropped non-zero constant factors. The product of the \(3j\)-symbols on the right-hand-side is non-vanishing if and only if (\S~34 of \cite{DLMF})
\be\label{eq:L-constraints}\begin{aligned}
    2 + L \quad\text{is even} \eqsp 0 \leq L \leq 2 \eqsp M = m + m' \, .
\end{aligned}\ee
These conditions are satisfied if and only if \(L = 0\) or \(L=2\) (we do not need the conditions on \(M\) for our argument). 
The fact that $F$ has a non-zero $L=2$ mode shows that spatial translations are not a subalgebra, because $F$ is proportional to 
\be
    \lb[ 3+s - \tfrac{1}{2} L(L+1) \rb] = 
    \begin{cases}
        3+s \neq 0 & \text{for } L = 0 \\
        s & \text{for } L = 2
    \end{cases}
\ee
(recall that for the class of spacetimes considered in this paper \( 0 \leq s < 1\)).
Only when \(s = 0\), does $F$ not have a \(L=2\) mode. 

In summary, for the usual BMS algebra \(\mf{bms} \cong \b_{s=0}\), we see that in all three cases \(F\) is a spherical harmonic with \(L = 0,1\) and hence $F$ is a translation. Therefore, when \(s=0\) the translation subalgebra is preserved under the Lie bracket of \(\mf{bms}\),  i.e. there is a preferred \(4\)-dimensional Lie ideal of translations in the BMS algebra. However, when \(s\neq 0 \), the translations \(f\), in general, change by \(F\), which contains a \(L=2\) spherical harmonic. As a result, translations are not preserved by the Lie bracket of \(\b_s\) and are not a preferred subalgebra (Lie ideal) of \(\b_s\). Case 1 also demonstrates that there is no smaller subalgebra of ``time translations'', i.e., spherical harmonics with $\ell=0$.
In fact, the above argument can be generalized to show that there is no finite-dimensional Lie ideal of \(\b_s\) when \(s \neq 0\).

\section{Spatially open FLRW spacetimes}
\label{sec:open}

In this appendix, we show that the techniques developed in this paper do not (straightforwardly) apply to FLRW spacetimes with spatially open slices, except for the special case of the Milne universe which is asymptotically flat. The calculations are similar to those in \cref{sec:conformalcompletionFLRW}. Apart from the Milne universe, which is a vacuum solution, we will only consider the cases with equation of state parameter \(w \geq -1/3\), since such spacetimes have a null boundary at infinity \cite{Harada:2018ikn}.\\

The physical metric $\hat{g}_{ab}$ of decelerating, spatially open FLRW spacetimes is described by the line element
\begin{equation}\label{eq:FLRW-open}
	d\hat s^2 = a^2(\eta) \left( - d\eta^2 + dr^2 + \sinh^2 r \; S_{AB} dx^A dx^B \right),
\end{equation}
with $r\in \left[ 0, \infty \right)$ and \(\eta < \infty\). The lower bound of \(\eta\) depends on the behaviour of the scale factor for each case, but we will not consider this in detail as we are concerned with the asymptotics at infinity.
First, we perform a coordinate transformation to $(U,V, x^A)$ coordinates, with
\be
\eta = \arctanh \left(\tan \tfrac{V}{2} \right) +  \arctanh\left(\tan \tfrac{U}{2} \right) \eqsp r = \arctanh \left(\tan \tfrac{V}{2} \right) - \arctanh\left(\tan \tfrac{U}{2} \right)
\ee
and $U, V < \pi/2$. By making a conformal transformation with conformal factor
\begin{equation}
	\Omega = \left(\cos U \; \cos V\right)^{\frac{1}{2}}  \; a^{-1}(U,V) \; ,
\end{equation}
the conformally rescaled line-element is 
\begin{equation}
	d s^2 = \Omega^2 d\hat s^2 = - d U d V + \sin^2 \tfrac{U-V}{2} \; S_{AB} dx^A dx^B \; .
\end{equation}
Since the metric is smooth everywhere including at the boundary with $V=\pi/2$, we can attach this boundary \(\scri\) to our spacetime. In each of the cases considered below \(\Omega \hateq 0 \) and \(\grad_a \Omega \not\hateq 0\). Thus, the surface \(\scri\) is future null infinity and it is clear from the form of the metric that $\scri$ is null.

The smoothness of the conformal factor near $\scri$ depends on the scale factor \(a\). As we will see this behaviour is rather different from the conformal factor of FLRW spacetimes with spatially flat topology considered in the main paper, except for the case of the Milne universe.

\paragraph*{Case 1: Milne universe.} Consider first vacuum case i.e. \(\hat T_{ab} = 0\). The Friedmann equations then give us the scale factor
\be
    a(\eta) = e^\eta ,
\ee
where we have set \(a(\eta = 0) = 1\). Note that as \(V \to (\pi/2)^-\) we have \(\eta \to +\infty\) and \(e^{-\eta} \sim (\cos V)^\half \sim (\pi/2-V)^\half\) and so we can expand the conformal factor as
\be\label{eq:exp-eta-exp}
    \Omega = e^{-\eta} \lb( \cos U \cos V \rb)^\half \sim (\pi/2 -V) \lb[ 1 - \tfrac{1}{24} (\pi/2 - V)^2 + \ldots \rb]  ,
\ee
where we have suppressed smooth functions of \(U\). Note that the conformal factor is smooth and satisfies \(\Omega \hateq 0\) and \(\nabla_a \Omega \not\hateq 0\). Thus, the Milne universe can be analyzed using the standard methods for asymptotically flat spacetimes. This is not surprising given that this spacetime is merely a part of Minkowski spacetime, being the interior of the future light cone of a point in Minkowski spacetime.

\paragraph*{Case 2: \(w > - 1/3\).} For this case we have \(0 < s < 1\), and the scale factor is given by
\be
    a(\eta) = \left[\frac{\sinh \left(\frac{1-s }{s} \; \eta \right) }{\sinh \left( \frac{1-s}{s} \;  \eta_0 \right) } \right]^{s/(1-s)} .
\ee
For convenience we choose $\eta_0$ such that $\sinh \left( \frac{1-s }{s} \; \eta_0\right)= 2$. Then, we have the following asymptotic expansion for the scale factor
\be
\begin{aligned}
	a^{-1}(\eta) =  (2\sinh\tfrac{1-s}{s} \eta )^{-s/(1-s)} &= e^{-\eta}\lb[ 1- e^{- \tfrac{2(1-s)}{s} \eta} \rb]^{-s/(1-s)} \\
	&= e^{-\eta} + \tfrac{s}{1-s} (e^{-\eta})^{1+\tfrac{2(1-s)}{s}} + \ldots
\end{aligned}
\ee
So that at leading order, the conformal factor can be expanded just as in \cref{eq:exp-eta-exp} to give
\be
    e^{-\eta} \lb( \cos U \cos V \rb)^\half \sim (\pi/2 -V) \lb[ 1 - \tfrac{1}{24} (\pi/2 - V)^2 + \ldots \rb] ,
\ee
where, as before, we have suppressed smooth functions of \(U\). However, the subleading term gives
\be\begin{aligned}
	(e^{-\eta})^{1+\tfrac{2(1-s)}{s}} \lb( \cos U \cos V \rb)^\half \sim (\pi/2 - V)^{1/s} + \ldots 
\end{aligned}
\ee
Hence, \(\Omega\) at leading order is \((\pi/2-V)\) but can have subleading terms which fall off as \((\pi/2-V)^{1/s}\). As a result, \(\Omega \hateq 0\) and \(\nabla_a \Omega \not\hateq 0\) at \(\scri\), but \(\Omega\) is not smooth, unless \(1/s\) is integer. This non-smoothness cannot be ``cured'' by choosing any \(\Omega^p\) for some \(p\).

For completeness, we note that the stress-energy tensor to leading order scales as
\be\begin{aligned}
	&T_{UU} \sim (2\sinh \tfrac{1-s}{s}\eta)^{-2} \sim (\pi/2 - V)^{\tfrac{1-s}{s}} \\
	&T_{VV} \sim (2\sinh \tfrac{1-s}{s}\eta)^{-2} (\sec V)^2 \sim (\pi/2 - V)^{-3+ \frac{1}{s}} \\
	&T_{UV} \sim T_{AB} \sim (2\sinh \tfrac{1-s}{s} \eta)^{-2} \sec V \sim (\pi/2 - V)^{-2+ \frac{1}{s}} \; ,
\end{aligned}\ee
where we have excluded constant factors and smooth functions of \((U,\theta,\phi)\).\\

\paragraph*{Case 3: \(w = -1/3\).} In this case we have $s=1$ and the scale factor in \cref{eq:FLRW-open} is
\begin{equation}
  a(\eta) = e^{\sqrt{1+c} \; \eta} \; ,
\end{equation}
where we have set \(a(\eta = 0) = 1\) and \(c = \tfrac{8\pi}{3}\rho_0 \neq 0\) with \(\rho_0\) the matter density at the time \(\eta = 0\). Consequently, the conformal factor for this scenario is
\be
\Omega = (e^{-\eta})^{\sqrt{1+c}} \lb( \cos U \cos V \rb)^\half \; .
\ee
Using the expansion in \cref{eq:exp-eta-exp} it is clear that while \(\Omega \hateq 0\) and \(\grad_a \Omega \not\hateq 0\), \(\Omega\) is not smooth since \(c \neq 0\). Just as for the case $w > -1/3$, the non-smooth behavior of $\Omega$ cannot be ameliorated by a power-rescaling as in the case of spatially flat FLRW spacetimes.  
The stress-energy components scale near $\scri$ as
\be\begin{aligned}
	&T_{UU} \sim c (\sec U)^2 \\
	& T_{VV} \sim c (\sec V)^2  \sim (\pi/2 - V)^{-2} \\
	& T_{UV} \sim T_{AB} \sim c \sec U \sec V  \sim  (\pi/2 - V)^{-1}\; ,
\end{aligned}\ee
where we again excluded constant factors and smooth functions of \((U,\theta,\phi)\).\\

From the above analysis we see that --- except for the Milne universe which is asymptotically flat --- the spatially open FLRW universes behave very differently from the spatially flat case considered in this paper.
Consequently, FLRW spacetimes with negative spatial curvature do not belong to the class of spacetimes with a cosmological null asymptote as defined in \cref{def:dna}. One possible approach to study the class of spacetimes mimicked after the behavior of spatially open FLRW spacetimes would be to introduce a smooth $\tilde{\Omega}$ that is related to $\Omega$ through some more complicated function which not just a power of \(\Omega\). We have not explored this interesting possibility in detail.



\bibliographystyle{JHEP}
\bibliography{asymp-FLRW}      
\end{document}